IAPS | Institute for AI Policy and Strategy

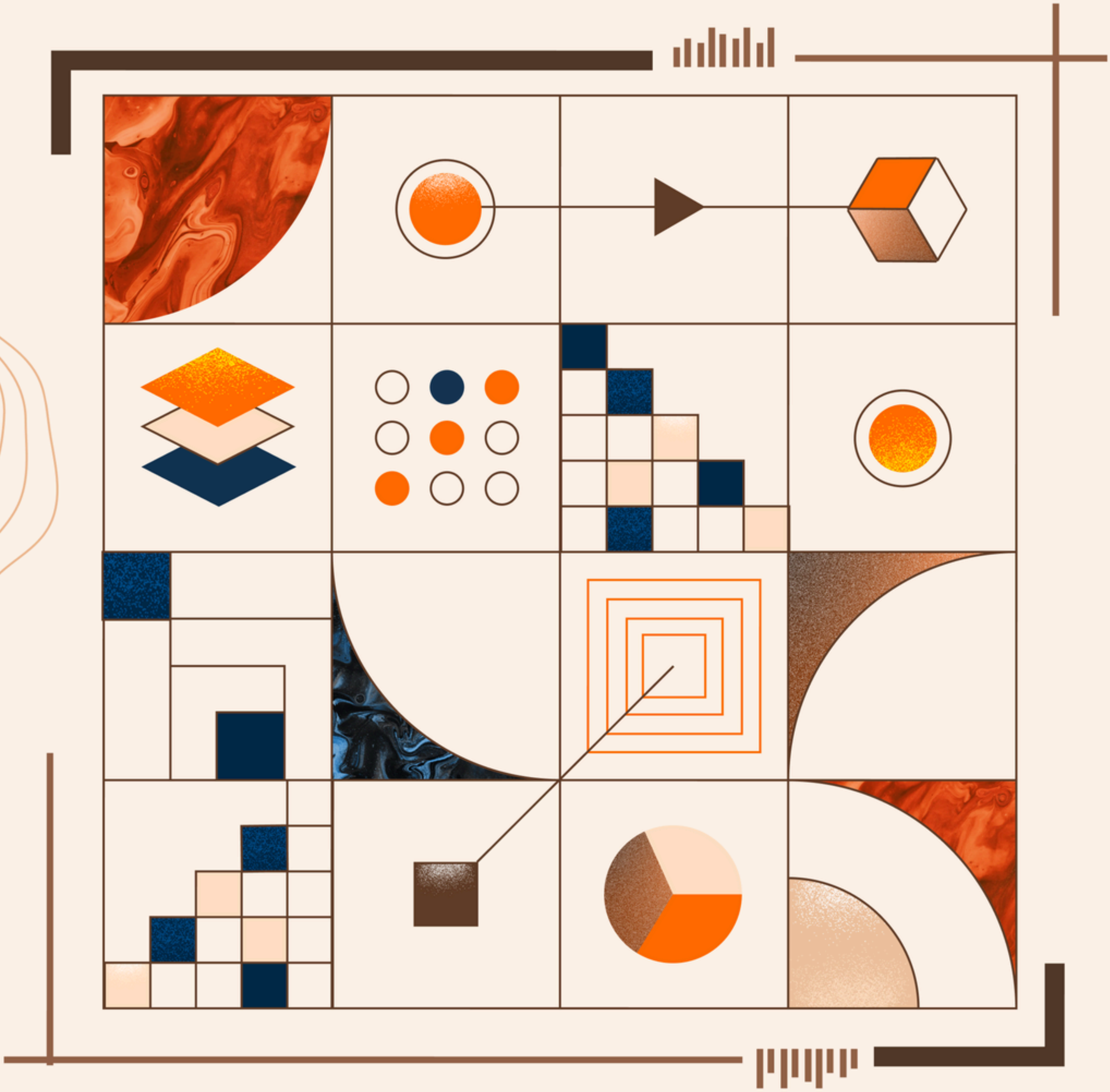

May 2025

# Expert Survey: AI Reliability & Security Research Priorities

By Joe O'Brien[1]*†, Jeremy Dolan[2]*, Jay Kim[3], Jonah Dykhuizen[2], Jeba Sania[4], Sebastian Becker[5], Jam Kraprayoon[1], and Cara Labrador[1]

* Equal contribution. † Corresponding author: joe@iaps.ai. [1] Institute for AI Policy and Strategy (IAPS).
[2] Independent Researcher. [3] Williams College. [4] Harvard Kennedy School. [5] Effektiv Spenden.

# Executive Summary

As AI systems accelerate toward broadly human-level performance amid unprecedented investment, reliability and security research is urgently needed to prevent severe harms and ensure AI can be deployed safely and reliably to realize its benefits. Progress is partly bottlenecked by uncertainty about which technical research directions offer the greatest risk reduction per marginal dollar.

To clarify priorities, **we surveyed 53 experts from academia, industry, and civil society. These experts rated subsets from a list of 105 technical areas** of AI reliability and security on *importance* (likelihood to significantly reduce AI harms) and *tractability* (ability for $10M USD to make significant progress) on a 1-5 scale. Due to the breadth of the list of research areas, respondents were encouraged to only rate areas they had personal expertise in, rather than the full list.

## Highest-Ranking Research Areas

| Research sub-area[1] | Importance (n) | Tractability (n) | Promise[2] |
|---|---|---|---|
| Emergence and task-specific scaling patterns | 5.00 (3) | 4.25 (4) | 21.25 |
| CBRN (Chemical, Biological, Radiological, and Nuclear) evaluations | 4.67 (3) | 4.33 (3) | 20.22 |
| Evaluating deception, scheming, situational awareness, & persuasion | 4.75 (4) | 4.25 (4) | 20.19 |
| Oversight and monitoring of LLM-agents | 4.67 (9) | 4.22 (9) | 19.70 |
| Cyber-capability evaluations | 4.50 (4) | 4.25 (4) | 19.13 |
| *… see full results in Appendix C.* | | | |

**The highest-ranked approaches emphasized preparing for emerging risks, with a strong focus on practical evaluation and monitoring over theoretical work:** six of the top ten most promising approaches center on improving evaluations of dangerous capabilities, while the top-ranked approach focuses on capability forecasting. Additionally, research on multi-agent interactions consistently ranked highly, suggesting growing concern that advanced multi-agent systems and their interactions present a new vector of risk.

[1] For the full list of research sub-areas including descriptions, see Appendix A.
[2] Promise = Importance × Tractability (max = 25)

**In addition to these highest-ranked sub-areas, eight sub-areas were rated as *highly important* but not *tractable* at a two-year, $10M USD scale.** All are either (i) applied security engineering (access control, supply-chain integrity, weight-key management, confidential compute) or (ii) deep-model understanding (mechanistic reasoning, eliciting latent knowledge). These likely require multi-year, larger-scale R&D programs.

## Implications for Funders and Policymakers

| Horizon | Recommended Action | Rationale |
|---|---|---|
| 0-2 yrs (< $10 M) | Fund independent dangerous-capability evaluations, scalable oversight tools, and multi-agent testbeds. | High promise, strong expert consensus, undercapitalized relative to risk. |
| 2-5 yrs (> $10 M) | Launch coordinated initiatives in AI-specific cyber and infrastructure security (e.g., supply-chain integrity, confidential compute). | Rated critical but not tractable at small scale; cross-disciplinary and infrastructure-heavy. |
| Enablers | i. Targeted talent programs<br>ii. Subsidised access to frontier models (e.g., NAIRR pilot, regulatory sandboxes)<br>iii. Government or multilateral signalling of priority research gaps | Addresses skills bottleneck and coordination failures. |

**Caveats:** Modest sample size (n=53) and uneven coverage across sub-areas limit statistical confidence; results should be read as directional. Future iterations will aim to broaden participation and capture updates in priorities closer to real time.

**Our study reveals a consistent message from respondents: significant, actionable opportunities exist within technical AI reliability and security research.** 52 out of 53 respondents identified at least one research direction as both important and tractable (scoring ≥ 4.0 on both dimensions). We consider this broad optimism about accessible, actionable opportunities a strong signal for well-resourced stakeholders interested in funding technical AI reliability and security research.

# Table of Contents

# Introduction

## Background

Artificial intelligence (AI) is rapidly advancing in many domains (Sevilla et al. 2022; Sevilla 2024; Maslej et al. 2025). Language models have demonstrated remarkable potential to scale capabilities along with more data, compute, and more effective algorithmic techniques (Kaplan et al. 2020; Hoffmann et al. 2022; Samborska 2025), while experts predict they will continue to become increasingly capable of performing complex tasks and taking extended autonomous actions (Kwa et al. 2025). CEOs of leading AI companies have publicly suggested that AI systems could reach human-level performance across a wide range of domains within the next 2-5 years,[3] and these companies are investing billions of dollars to reach this milestone first. The potential economic and geopolitical advantages from such systems have ignited competition between global powers, particularly between the United States and China, creating a growing national priority and additional incentives to accelerate development.

While it is widely recognized that AI reliability and security research is essential for ensuring that these increasingly powerful systems remain safe, reliable, and beneficial (Bengio et al. 2025), the impending arrival of powerful, general-purpose AI creates an urgent need for such research to keep pace. To do so requires addressing a range of technical challenges, from AI-specific concerns (such as alignment, control, and interpretability) to high-stakes versions of traditional cybersecurity and infrastructure threats (such as model theft, adversarial attacks, and infrastructure vulnerabilities).

## Motivation

Despite widespread awareness of this urgent challenge (Bengio et al. 2025), significant gaps remain in coordinating and prioritizing technical AI reliability and security (AI R&S) research efforts. Progress has been limited by several bottlenecks:

- **Resources:** While funding for AI R&S has increased in recent years, it remains inadequate relative to the scale and urgency of the problem.
- **Expertise:** The field faces a shortage of researchers with the necessary technical skills.
- **Uncertainty:** There is considerable uncertainty about which research directions offer the most promise for risk reduction per unit of effort.

This survey aims to address the uncertainty bottleneck by providing an assessment of research priorities across the technical AI R&S landscape. Through a survey of researchers across industry, academia, and civil society, **we evaluate 105 areas of technical AI R&S research (grouped into 20 high-level research categories) along dimensions of *importance* and *tractability*.**

[3] "I think we're probably three to five years away [from AGI]" — Demis Hassabis, CEO of Google DeepMind (Kantrowitz 2025). "We are now confident we know how to build AGI" — Sam Altman, CEO of OpenAI (Altman 2025). "We are rapidly running out of truly convincing blockers, truly compelling reasons why this [AGI] will not happen in the next few years." — Dario Amodei, CEO of Anthropic (Fridman 2024).

[Appendix A](#) summarizes all 105 sub-areas (and their corresponding higher-level categories) included in the survey. Readers looking for additional context on the research landscape may wish to review this taxonomy before proceeding to the results and analysis.

While several high-quality descriptive overviews of the technical AI R&S landscape exist (Bengio et al. 2025; Reuel et al. 2025; Anwar et al. 2024; The Singapore Consensus on Global AI Safety Research Priorities, 2025) none of these attempt to explicitly rank research directions. Likewise, the few published expert surveys to date either cover governance practices (Schuett et al. 2023) or gauge general attitudes toward "AI safety" as a whole (B. Zhang and Dafoe 2019; Grace et al. 2024); they do not present a comparative, quantitative priority list of technical sub-areas. In the absence of such guidance, funders, policymakers, and researchers—especially new entrants to the field—risk allocating attention and resources toward the most visible labs or currently prevailing research paradigms, potentially crowding out less-publicized but higher-leverage work (cf. Merton 1968).

**This study is the first to explicitly quantify expert judgment across a comprehensive set of technical AI R&S topics (105 sub-areas in 20 categories) and to produce a data-driven ranking of their "promise" (importance × tractability)**. By examining where experts agree and where they diverge, we highlight the research directions most widely regarded as both critical and amenable to near-term progress.

## Related Work

This survey builds upon several recent efforts to map the technical AI R&S landscape. Taxonomies from Anwar, *et al.* (2024), Reuel & Bucknall *et al.* (2024), and the *AI Assurance Technology Market Report* (AIAT Report 2024) informed the structure of our survey. Although released after our survey was designed, the *International AI Safety Report* (Bengio et al. 2025) is another relevant contribution. This work also follows prior related research by IAPS, notably *Mapping Technical Safety Research at AI Companies* (Delaney, Guest, and Williams 2024). Methodologically, our survey draws inspiration from the Centre for the Governance of AI's *AGI Safety and Governance Survey* (Schuett et al. 2023), which similarly aggregated expert judgments to identify high-priority policy interventions.

## Scope of Results

The central result of this survey is a ranking of AI R&S research areas based on their overall "promise," defined as the product of experts' assessments of each area's importance and tractability. There are many ways one could define *importance* and *tractability*; **the definitions used in our survey were designed to assess the reduction in severe harms expected from an actionable level of funding ($10M USD).**

- "Importance" was defined as agreement with the statement: *"Resolving the core challenges of this sub-area and implementing the resulting solutions would significantly reduce the risk of severe harm (loss of >100 lives or >$10 billion in economic impact) from AI."* [4] This framing was chosen to provide concrete thresholds that capture consequential harms while remaining accessible to experts from diverse backgrounds.
- "Tractability" was defined as agreement with the statement: "*an additional targeted investment of approximately $10 million over the next two years would lead to significant, measurable advancements in addressing this issue.*" This dollar amount and time horizon were chosen as a practical scale for interventions by researchers, funding organizations, and policymakers facing near-term resource allocation decisions.

Any definition of these terms inherently affects the resulting assessment. For example, some research directions may be potentially high-impact, but unlikely to produce implementable results within two years; this framing will naturally disadvantage those directions. Similarly, others may require substantially larger monetary investments to overcome development and implementation barriers. While our core results must be interpreted in accordance with the definitions detailed above, we have also separately highlighted several challenging (i.e., high-impact, low-tractability) research directions, which may be of interest to funders with particularly large resource pools or who are willing to take long-term bets on important interventions (see Results).

[4] Defining importance via quantitative thresholds for severe harm (e.g., lives lost, major economic impact) is analogous to practices in other established risk management domains. For instance, nuclear safety regulation employs quantitative health objectives related to mortality risks as part of its safety goals for evaluating reactor accident risks (U.S. Nuclear Regulatory Commission 1986).

# Methods

## Survey Design

The primary objective of this survey was to **systematically capture experts' judgments of the importance and tractability of technical interventions related to AI reliability and security**, with particular attention to large language models (LLMs) and their underlying architectures. The survey was developed through an iterative process involving literature reviews, consultations with domain experts, and pilot testing to ensure clarity, relevance, and comprehensive coverage of significant technical interventions.

### Taxonomy

We developed a taxonomy of technical reliability and security research consisting of 105 sub-areas organized into 20 broader categories. The initial set of categories and sub-areas was drawn from recent literature, particularly Anwar et al. (2024), which was chosen for its structured overview of current technical challenges specific to LLM safety. This foundational source provided a well-defined categorization aligned with ongoing discussions within relevant research communities. The taxonomy used in the survey, with brief descriptions of all areas, is reproduced in Appendix A.

Items selected for inclusion met two primary criteria:

1. **Technical Focus:** To maintain the scope of the survey on a single mode of intervention, only research tracks with a "technical" focus were included. Broader governance and policies work, as well as investigations of sociotechnical factors, were intentionally omitted.
2. **Interventions on or around the model:** Interventions with a direct effect on model behavior, evaluation of model behavior, model design, or a model's immediate environment (including the physical infrastructure on which it runs, as well as tooling, monitoring, and access control systems) were prioritized over more distal reliability and security remedies (e.g., policy and governance).

This focus on technical and model-centric interventions aligns with the survey's goal of assessing tractability and importance specifically within technical AI R&S. This narrowed scope allows us to develop a more detailed understanding of technical intervention points and lays the groundwork for future surveys to explore complementary areas, such as AI governance and sociotechnical interventions.

### Expert Consultations

The initial set of items underwent significant refinement based on consultations with approximately 15 experts specializing in machine learning and AI R&S. Experts were chosen based on their publication records, academic credentials, or industry experience in relevant research domains. These consultations occurred over a period of approximately three weeks between late October and November 2024, involving structured one-on-one discussions aimed at verifying item relevance, clarity, and completeness.

During consultations, experts suggested additions, removals, or revisions to survey items, informed by their research experience and judgment regarding practical tractability and importance. This refinement process involved subjective judgments by the authors, who aimed to balance comprehensive coverage with survey manageability, thereby potentially introducing selection biases. Nonetheless, we sought in our decisions on survey design to reflect the consensus expert opinion regarding important and distinct approaches to reflect in the survey taxonomy.

### Survey Piloting

We conducted a pilot phase in December 2024 prior to the survey's official launch, involving one-on-one interactions with a small group of experts. The pilot phase assessed the interpretability of questions, definitions of "importance" and "tractability," and overall survey usability. Pilot feedback led to specific adjustments, notably improvements in clarity of rating scales, item descriptions, and clarity of framing text.

## Survey Administration

The finalized survey was administered online through the Qualtrics platform from December 21, 2024 to March 4, 2025. Responses were collected anonymously. Respondents were first prompted to select one or more high-level categories that aligned with their expertise from the presented list of 20 categories (e.g., "Scalable oversight and alignment techniques," "Robustness," "Cybersecurity for AI models"; see Appendix A for more). For each category chosen, respondents were then shown a series of sub-areas (e.g., a sub-area of "Scalable oversight and alignment techniques" is "Iterated Distillation and Amplification").

For each of the sub-areas in a chosen category, respondents were provided a brief description (reproduced in Appendix A) and asked a series of questions. Respondents then assessed that sub-area along dimensions of importance and tractability.

- **Importance:** "Resolving the core challenges of this sub-area and implementing the resulting solutions would significantly reduce the risk of severe harm (loss of >100 lives or >$10 billion in economic impact) from AI."

- **Tractability:** "An additional targeted investment of approximately $10 million over the next two years would lead to significant, measurable advancements in addressing this sub-area's underlying challenges."

Respondents were asked to indicate their level of agreement with the statement based on a 5-point Likert scale: "Strongly Disagree," "Disagree," "Neutral," "Agree," "Strongly Agree." Respondents also had the option to say "I don't know," and were encouraged to skip any sub-areas they were insufficiently familiar with. Each sub-area also contained an open-ended question asking respondents what indicators would best measure progress in this area and for suggestions to improve work in the area.

After assessing each sub-area in a category, respondents were also asked three free-response questions on what they saw as the most impactful challenge and key obstacles to progress in their chosen area, and any sub-areas missing from the survey.

Full instructions from the survey are reproduced in Appendix B.

## Sample

The expert sample for this survey was recruited through a structured, multi-pronged approach designed to identify individuals with demonstrable research contributions or recognized expertise relevant to the 105 technical AI R&S sub-areas detailed in our comprehensive taxonomy (see Appendix A).

The primary component of our recruitment strategy involved systematically identifying and inviting the first authors of the key publications cited as exemplars for each specific sub-area within our taxonomy, as well as secondary authors with relevant domain expertise and level of experience. This ensured that our initial outreach was grounded in established research contributions directly aligned with the survey's content.

To complement this and enhance the breadth of expertise, this initial list of potential participants was augmented through targeted outreach. This involved:

- Identifying additional relevant experts through supplementary, focused literature reviews within specific domains of the taxonomy.
- Consultations with subject matter experts to identify leading figures or individuals with unique perspectives in niche or emerging sub-areas, ensuring a more comprehensive inclusion of specialized knowledge.

In total, 515 researchers were invited to participate in the survey, and 53 completed an assessment of one or more sub-area (mean: 5.2, SD: 3.57), totaling 546 assessments across all sub-areas. This was a response rate of 10.6%, roughly comparable to similar surveys of AI experts (such as Grace et al. [2024], which garnered a 15% response rate).

470 researchers were initially contacted on December 21, 2024, with reminders sent on January 7, January 15, and January 21, 2025. After preliminary analysis of responses by sub-area, 45 additional experts for categories underrepresented in the initial data were contacted on February 22, with reminders sent on March 3. All responses occurred between December 21, 2024 and March 4, 2025.

## Data Processing

Sub-areas that received two or fewer ratings in either importance or tractability were excluded from quantitative analysis due to insufficient data for meaningful statistical interpretation. Exclusions are listed in Appendix C and limitations related to this exclusion are discussed in the Limitations section below.

For each sub-area with sufficient responses, we calculated mean scores for importance and tractability. We then computed a “promise score” for each sub-area by multiplying these means, reflecting our assessment that areas must excel in both dimensions to warrant prioritization. Standard deviations were calculated to assess consensus levels among respondents.

## Demographics

43 of the 53 respondents (81%) completed the optional demographic survey. Roughly half of respondents were from academia (PhD students, postdoctoral researchers, or faculty/professors), with the other half primarily affiliated with industry or non-profit organizations. Only one respondent identified their primary affiliation as governmental, while three identified their current position as “policymaker or other professional with equivalent experience.” Reflecting our survey’s focus on technical research, 84% of respondents identified their primary area of expertise as “machine learning/AI” or “computer science/engineering,” with only 7% listing “public policy/governance.”

For full demographic results, see Appendix E.

# Results

## Most Promising Sub-Areas

We identified the most promising research directions based on the product of each sub-area's average *importance* and *tractability* ratings, resulting in what we term a "promise score." To improve robustness in the results, we only include approaches for which at least three respondents provided ratings for both *importance* and *tractability* (this filtered out 29 sub-areas, listed in Appendix D). Finally, we filtered for areas that achieved an average rating of at least 4 ("Agree," on our 5-point Likert scale) for both *importance* and *tractability*. This resulted in the following 15 most promising research directions. A description of each sub-area (as shown to respondents during the survey) is given in Appendix A. Full results, as well as additional data on areas of highest and lowest consensus, are contained in Appendix C.

| | Research sub-area | Importance | (n) | Tractability | (n) | Promise |
|---|---|---|---|---|---|---|
| **1** | Emergence and task-specific scaling patterns | 5 | (3) | 4.25 | (4) | **21.25** |
| **2** | CBRN (Chemical, Biological, Radiological, and Nuclear) evaluations | 4.67 | (3) | 4.33 | (3) | **20.22** |
| **3** | Evaluating deception, scheming, situational awareness, and persuasion | 4.75 | (4) | 4.25 | (4) | **20.19** |
| **4** | Oversight and monitoring of LLM-agents | 4.67 | (9) | 4.22 | (9) | **19.7** |
| **5** | Cyber evaluations | 4.5 | (4) | 4.25 | (4) | **19.13** |
| **6** | Detecting and addressing previously unmeasured or latent capabilities | 4.38 | (16) | 4.25 | (16) | **18.59** |
| **7** | Multi-agent metrics and evaluations | 4.43 | (7) | 4.14 | (7) | **18.35** |
| **8** | Multi-agent security | 4.57 | (7) | 4 | (6) | **18.29** |
| **9** | Quantifying cyber threats from advanced capabilities | 4.25 | (4) | 4.25 | (4) | **18.06** |
| **10** | Manage conflicts between different values | 4.5 | (4) | 4 | (4) | **18** |
| **11** | Safety and emergent functionality in multi-agent interactions | 4.57 | (7) | 3.86 | (7) | **17.63** |
| **12** | Mechanistic understanding and limits of LLM reasoning | 5 | (3) | 3.5 | (4) | **17.5** |
| **13** | Validating and applying interpretability methods | 4.38 | (8) | 4 | (7) | **17.5** |
| **14** | Evaluation methodology and metrics | 4.14 | (14) | 4.2 | (15) | **17.4** |
| **15** | Control mechanisms for untrusted models | 4 | (4) | 4.25 | (4) | **17** |

## Strategic Long-Term Opportunities

Several research areas emerged with strong consensus on their importance despite lower perceived tractability. These critical domains—primarily in security implementation and applied systems—represent strategic priorities that may require longer timelines and substantial resource commitments to achieve meaningful progress. While they present significant technical challenges, their high importance scores (all ≥4.0) indicate these areas warrant further consideration, despite apparent implementation barriers.

The following table shows research areas with the largest positive gaps between importance and tractability:

| | Research sub-area | Importance | (n) | Tractability | (n) | Gap |
|---|---|---|---|---|---|---|
| **1** | Access control and interface hardening | 4.75 | (4) | 2.75 | (4) | **2.00** |
| **2** | Supply chain integrity and secure development | 4.57 | (7) | 3.00 | (6) | **1.57** |
| **3** | Mechanistic understanding and limits of LLM reasoning | 5.00 | (3) | 3.50 | (4) | **1.50** |
| **4** | Preventing model self-exfiltration | 4.00 | (4) | 2.75 | (4) | **1.25** |
| **5** | Weight security and key management | 4.50 | (4) | 3.25 | (4) | **1.25** |
| **6** | Confidential computing and environment isolation | 4.57 | (7) | 3.43 | (7) | **1.14** |
| **7** | Detecting modified models or poisoned data | 4.00 | (4) | 3.00 | (4) | **1.00** |
| **8** | Eliciting Latent Knowledge (ELK) | 4.25 | (8) | 3.25 | (8) | **1.00** |
| **9** | Iterated Distillation and Amplification (IDA) | 4.00 | (9) | 3.22 | (9) | **0.78** |

# Discussion

## Key Findings

### Patterns Among Top-Ranked Approaches

Analysis of the highest-ranked approaches reveals several patterns that should inform research prioritization:

**Anticipating Dangerous Capabilities**

Eight of the top ten approaches focus on ensuring society is not caught unprepared by sudden advancements in harmful AI capabilities:

- Six approaches center on improving evaluations of dangerous capabilities.
- The top-ranked approach "Emergence and Task-Specific Scaling" (T=4.25, I=5.0) focuses on formalizing and forecasting new capabilities as models scale.
- The fourth-ranked approach "Oversight and monitoring of LLM-agents" (T=4.22, I=4.7) addresses tracking agent actions.

This corroborates findings from previous expert surveys that dangerous capability evaluations are among the most promising directions for further research (Schuett et al., 2023). Of the 31 respondents who prioritized these approaches, five explicitly noted in free-form responses that lack of funding for top talent was the main bottleneck to progress. While dangerous capability evaluations are currently being researched at frontier AI companies and by a small number of third-party bodies (such as ScaleAI, METR, Apollo Research, and government initiatives like the US AI Safety Institute and the UK AI Security Institute), our findings suggest this direction is still undercapitalized relative to its importance.

**Multi-Agent Interactions Emerge as a Critical Area**

All sub-areas of the *Multi-Agent Interactions* category scored within the top 30 (out of 105 total), a feat matched only by *Domain-specific Evaluations*. Multi-agent metrics and evaluations (T = 4.14, I = 4.43)[5] and multi-agent security (T = 4.14, I = 4.57) ranked 7th and 8th highest, respectively. This highlights growing concern that advanced multi-agent systems present novel risks distinct from those posed by single agents (Hammond et al. 2025).

**Practical Evaluation Over Theoretical Frameworks**

A clear divide emerged between approaches focused on practical evaluation and those

[5] *T* refers to the mean of respondent answers to sub-area tractability, and *I* refers to the mean of respondent answers to sub-area importance. We will use this shorthand throughout the paper.

centered on theoretical frameworks:

- Evaluation-focused approaches dominate the top rankings. Nine of the top 15 approaches emphasize evaluation, detection, or monitoring of harms (such as "Emergence and task-specific scaling patterns," "CBRN evaluations," and "Evaluating deception and scheming”), rather than solutions aimed at root causes, or theoretical research.
- Theoretical approaches clustered at the bottom. Traditional machine learning theory (such as "Double descent and overparameterization," "Implicit bias of optimization algorithms," and "Optimization and loss landscape analysis") received lower rankings. These approaches address foundational questions about the nature of modern AI systems but were evaluated as less directly impactful for addressing large-scale concerns.
- A preference for practical interventions is also evident in the high rankings of approaches targeting specific risk domains like CBRN, cyber, and deception.

## Additional Patterns

**Implementation Challenges (High Importance, Low Tractability)**

Analysis of essential but challenging approaches reveals several patterns:

1. Security implementation challenges: Six of the ten largest importance-tractability gaps come from cybersecurity or infrastructure security domains.
2. Implementation over theory: These gaps primarily appear in applied implementation areas rather than theoretical frameworks.
3. Technical complexity: Many of these approaches require complex technical solutions spanning multiple domains (e.g., hardware, software security, ML systems).

**Consensus Areas (Low Variance)**

Areas with the strongest consensus among experts share several characteristics:

1. Evaluation focus: All five areas with the lowest variance relate to evaluation, monitoring, or detection approaches.
2. Strong promise: Areas with the strongest consensus frequently appear among the highest promising scores (four of the five appear in the top 15).
3. Concrete objectives: These approaches tend to have well-defined objectives and clear implementation paths.
4. Applied methods: Most low-variance areas focus on practical applications rather than theoretical frameworks.

**Areas of Disagreement (High Variance)**

Areas with the greatest expert disagreement reveal several interesting patterns:

1. Implementation uncertainty: The highest variance appears in areas focused on implementing safety mechanisms rather than evaluating risks.
2. Technical complexity: Many high-variance areas involve complex technical challenges spanning multiple domains.
3. Hardware and infrastructure security: Three hardware/infrastructure security topics appear in the top 10 highest variance list, suggesting some fundamental disagreement on effective paths forward for this domain.
4. Theoretical foundations: Several foundational research areas show high variance, particularly in tractability assessments.

## Four Example Promising Sub-Areas

Below we discuss four illustrative promising sub-areas in detail. For each, we provide the description given during the survey, why research in this area matters for preventing severe harms, ratings according to the survey, exemplary research and funding, and key respondent insights. The discussions provide example starting points for funders and others who want to explore individual sub-areas further.[6] We chose these particular sub-areas as they are either highly ranked or received the highest number of responses, compared to the other top 15 sub-areas.

### Emergence and Task-Specific Scaling Patterns

**Description:** Formalizing and forecasting the emergence of new capabilities as models scale, investigating whether scaling alone can produce certain capabilities, and designing methods for discovering task-specific scaling laws.

**Ranking:** Highest ranked sub-area according to promise score (T = 4.25, I = 5).

**Why It Matters:** Anticipating and mitigating potential severe harms from future AI capabilities necessitates accurate forecasting of when and how these capabilities may emerge. This foresight enables the proactive implementation of safety measures and the formulation of informed policy guidelines, ensuring that safety responses do not lag behind the development of capabilities. The phenomenon of emergence—where capabilities such as multi-step reasoning and instruction following arise unexpectedly with scaling (Wei et al. 2022)—is particularly pertinent, as it is inherently surprising and challenging to predict. Employing task-specific scaling laws facilitates more granular predictions, enhancing our ability to anticipate and address emerging capabilities effectively (Caballero et al. 2023).

[6] Additional sources to further ground prioritization of AI R&S research and funding include Bengio et al. (2025), The Singapore Consensus on Global AI Safety Research Priorities (2025) and Fist (2025) (particularly the sections on “basic science,” “security,” and “evidence of risks”).

**Exemplary Research and Funding:** Example work has been conducted across academic institutions (Caballero et al. 2023), AI companies (Ganguli et al. 2022), and research non-profits (Sevilla 2024). Ruan, Maddison, and Hashimoto (2024), from Stanford and the University of Toronto, introduce the concept of *observational scaling laws*, showing that small model performance on specific benchmark tasks can anticipate the emergence of qualitatively new behaviors in larger models. Jones et al. (2025) at Anthropic propose a methodology for forecasting model failures at deployment scale.

This line of research has also attracted philanthropic support: EpochAI, a non-profit research institute working on forecasting AI capabilities, has received funding from Open Philanthropy (OP) and Jaan Tallinn (Epoch AI 2025). Furthermore, OP and Schmidt Futures supported work on "inverse scaling," focused on cases of worse performance with increased scale (McKenzie et al. 2024). Finally, OP's April 2025 RFP on "Improving Capability Evaluations," is, among other things, interested in research exploring how existing measures can predict emergence of new capabilities (Open Philanthropy 2025a).

**Respondent Insights:** One respondent worried that the safety benefits from this area of research could be offset by advances in dual-use AI capabilities that it would enable. They suggested future support would be best targeted toward enabling independent researchers or academic labs to work with cutting-edge models.

## Oversight and Monitoring of LLM-agents

**Survey Description:** Building automated oversight and monitoring tools to track LLM-agent actions.

**Ranking:** 4th highest rated sub-area (T = 4.22, I = 4.7), with the 3rd highest number of responses (9).

**Why It Matters:** LLM agents can autonomously execute multi-step tasks—such as writing code, conducting research, or interacting with external systems—without continuous human supervision. This autonomy increases the risk of unintended or malicious actions, including data breaches, financial fraud, or physical harm. Without monitoring, failures could go undetected until significant damage occurs. As agents are integrated into critical domains like finance, healthcare, and cybersecurity, robust oversight mechanisms are essential to ensure accountability, enforce constraints on agents, and prevent large-scale harm.

**Exemplary Research and Funding:** There is ongoing academic research which proposes different frameworks for monitoring and evaluating LLM agents (T. Yuan et al. 2024; Ruan et al. 2024; Guo et al. 2024). A relevant field, which in our taxonomy belongs to the sub-area "Control mechanisms for untrusted models," but has important overlap, is "AI control." AI control is a category of plans which tries to ensure the safety of AI systems even if they try to avoid control. This work, developed at the non-profit research lab Redwood Research, initially focused on non-agentic LLMs (Greenblatt et al. 2024), but is being extended to agents, too (Bhatt et al. 2025). The work at Redwood Research has been made possible by several philanthropic funders, such as OP (Open Philanthropy 2023), the Survival and Flourishing Fund (Survival & Flourishing Fund 2024) and The Future of Life Institute (Future of Life Institute 2023). AI control has been endorsed by Google DeepMind (Shah et al. 2025) and the UK AI Security Institute staff (Tomek Korbak et al. 2025).

**Respondent Insights:** Respondents suggested that this area would strongly benefit from additional funding, but that because of overlap with rote surveillance, it would be important to prioritize funding directions that would institute monitoring regimes outside of direct government control.

## Evaluation Methodology and Metrics

**Survey Description:** Research focuses on designing holistic, theory-grounded metrics (e.g., focused on more than just harmlessness), accounting for scaffolding in evaluations, and characterizing safety–performance trade-offs.

**Ranking:** 12th highest rated sub-area (T = 4.2, I = 4.14), with the second highest number of responses (14) and the second lowest variance, suggesting a uniquely widespread optimism compared to most other sub-areas found in this survey.

**Why It Matters:** Research into holistic evaluation methodologies and statistically sound metrics is essential for ensuring that we are accurately measuring what matters, and testing models in ways that reflect the full complexity of real-world use. Traditional benchmarks often emphasize narrow metrics like accuracy on static datasets, but this can give a false sense of capabilities. In practice, models may perform well on these curated test sets yet fail in interactive, adversarial, or open-ended contexts. Denser evaluations which consider safety–performance trade-offs, trustworthiness, and robustness in addition to test set accuracy can expose blind spots in current systems and improve our ability to forecast model behavior in deployment settings. Success would enable more transparent, accountable, and adaptive assessments of AI systems, helping to prevent the deployment of models whose safety properties are poorly understood or misrepresented.

**Exemplary Research and Funding:** Evaluations as a tool to measure the capabilities and tendencies of AI systems receive substantial attention from AI companies (Ziv and Anadkat 2024), non-profit labs (e.g., METR), government institutes (UK AI Security Institute 2024) and academia. Safety–performance trade-offs have been subject to research at AI companies (Bai et al. 2022). OP has disbursed grants worth $25 million on “benchmarking LLM agents on consequential real-world tasks” (Open Philanthropy 2024), the need for which had been discussed in Kiela et al. (2021).

**Respondent Insights:** Respondents emphasized the broad range of opportunities for making "measurable progress" in evaluation methodology and metrics. Several highlighted the potential for importing methodological best practices from experimental psychology, such as validity testing and control of confounding variables. Others pointed to a large and urgent need for better automation and tooling to support scalable, rigorous assessments. The area's consistent, high ratings for importance and tractability were reflected in many comments, with one respondent concluding: "Anything that can bring clarity and focus to the science of evaluations will be enormously helpful."

## Detecting and Addressing Previously Unmeasured or Latent Capabilities

**Survey Description:** Developing strategies to uncover latent harmful abilities within AI models and prevent models from exhibiting undesirable behaviors such as "sandbagging" or deceptively underperforming during evaluations.

**Ranking:** This approach garnered the highest number of responses from respondents out of all approaches (16, 30% of respondents) and was ranked 7th according to the promising score.

**Why It Matters:** Harmful capabilities include deception, persuasion, and the self-extraction of a model's own capabilities. AI models can become deceptive if it serves their goals (Ward et al. 2023) and even hide their deceptive behaviour to the user (Scheurer, Balesni, and Hobbhahn 2024). Deception is in itself harmful as it causes users to have false beliefs. "Sandbagging," the tendency for a model to underperform during formal evaluations of dangerous capabilities (van der Weij et al. 2025), can have even more severe consequences. The concern is that a model might be engineered—or may evolve tendencies—such that it exhibits low performance on safety-critical benchmarks while in reality possessing the latent capacity to engage in harmful capabilities, which will then unfold after deployment.

**Exemplary Research and Funding:** Some of the relevant work on deception was produced at academic institutions (Ward et al. 2023) and non-profit research organizations supported by philanthropic funders (Scheurer, Balesni, and Hobbhahn 2024) and the ML Alignment & Theory Scholars Program (Weij et al. 2025). In an RFP closed in April 2025, OP is eliciting proposals on "Experiments on alignment faking" and "Evaluating whether models can hide dangerous behaviors," 2 of 21 sub-areas on which they want to spend $40 million (Open Philanthropy 2025b). What seems particularly relevant from a funder's perspective in this area is that developers can have an incentive to understate a model's capabilities to meet regulatory standards (van der Weij et al. 2025). This indicates structural reasons why detecting and measuring latent harmful capabilities might not be sufficiently invested in by the major AI companies. It should be noted, however, that companies have been doing some work in this area, e.g., Anthropic (Järviniemi and Hubinger 2024).

**Respondent Insights:** Respondents suggested that red-blue team exercises and the development of model organisms could be useful tools for uncovering latent harmful capabilities. One concern expressed was that while it may be feasible to detect capabilities researchers already suspect might exist, identifying unexpected or unanticipated abilities could be significantly more challenging.

## Limitations

As a first iteration, our research has several limitations that should inform the interpretation of its results:

### Response Count

Despite extensive outreach, we received responses from a total of 53 experts—below our initial goal. This sample size limits the survey's reliability as a gauge for the perspectives of the broader AI R&S expert community. Accordingly, we encourage readers to consider this survey as one piece of evidence among many, rather than as ground truth. For future iterations, we may consider offering an honorarium or other incentives to encourage greater participation (Grace et al. 2024).

### Response Distribution

Several categories and sub-areas consistently received fewer responses, limiting our ability to provide robust quantitative insights across all areas. Although the distribution of outreach targeting subject matter experts was approximately balanced across categories, this balance did not carry over into responses. The anonymous nature of the survey prevents confirming if experts were biased toward their own research areas (though we did ask experts to answer questions specifically in areas aligning with their expertise, in an attempt to reduce respondent bias by making it more uniform across areas). The concentration of machine learning academics may have skewed responses toward popular topics in that field while potentially underrepresenting views on areas like privacy or fairness. Other possible explanations include network bias, or organic forwarding of the survey to experts not formally recruited.

### Design Choices for Scope of Impact

Our specific definitions of *impact*, *tractability*, and *importance*—including number of deaths prevented, time horizons, and funding amounts—may have disadvantaged certain research directions, especially long-term approaches requiring more than $10M USD in investment. However, we defined these specific parameters to ensure the results provided actionable insights for funders, researchers, and policymakers, with the objective of conveying the value of near-term marginal contributions (as opposed to "grand-challenge"-style bets).

### Evolving Research Landscape

This survey captures expert perspectives during a specific period. New reliability and security challenges have already emerged since this survey concluded in early 2025, including challenges related to reasoning models and inference-time compute scaling following releases like Deepseek's R1 models (DeepSeek-AI et al. 2025). A "live," continuously updated future version of the survey could track real-time changes and update categories accordingly, and we invite interested readers to contact us to explore the feasibility of such a project.

## Policy Implications

Policy can be a powerful instrument in mitigating critical research gaps. Specifically, government can directly fund neglected areas, incentivize investment, play a coordination role, strengthen AI talent pipelines, and expand researcher access to frontier models.

### Direct Funding

Government's role in promoting research will depend on the source of the research gap. In some instances, research areas may be underfunded because of limited financial resources. To address this, Congress and relevant executive agencies, such as the National Science Foundation, should consider directly appropriating funding toward the most promising research areas identified in this report. Respondents classified areas as tractable if a $10M USD investment over the next two years would yield substantial improvements in AI R&S—a modest sum by government grantmaking standards that could yield significant dividends in terms of increased reliability leading to broader adoption. In addition, government is uniquely positioned to direct higher sums to underfunded research areas that score high on importance but low on tractability (e.g., supply chain security, access control and interface hardening). The AI R&D ecosystem may be unable to address these gaps without large, long-term investments usually provided by government.

### Incentivizing investment

Short of providing direct funding, government agencies can use various indirect methods to incentivize additional investment in promising AI R&S research areas. These may include:

- **Listing identified research gaps as official government priorities**, thereby signaling their interest and spurring investment from industry, academia, and civil society.
- **Promoting AI R&S research** through tax incentives or subsidies.
- **Incorporating specific AI R&S research commitments** into broader agreements with industry stakeholders.
- **Lowering the cost of research** by leveraging existing frameworks such as the National Artificial Intelligence Research Resource (NAIRR) pilot, which provides computational, data, software, model, training and user support resources.

### Coordinating Investment

In other instances, research areas may be neglected not because of a lack of resources but because of limited awareness among funders that specific funding gaps exist. This represents a coordination problem that government can help alleviate by identifying and proactively communicating research needs internally and externally, targeting relevant government researchers, industry stakeholders, academia, and the broader research community. As part of its efforts to identify future research gaps in real time, government could establish public-private information-sharing mechanisms, giving it not only a comprehensive view of domestic AI R&D but also insights into progress toward advanced AI thresholds within AI companies. This coordination role may extend past the research stage, with government officials providing incentives for AI developers and deployers to incorporate best practices learned from AI R&S research.

### Strengthening Talent Pipelines

Underinvestment may also stem from the limited availability of specialized talent to conduct high-quality research. In the short term, where organizations lack the required talent to conduct relevant research, government can sponsor upskilling efforts such as scholarships for individuals to study specific critical AI R&S areas and develop specialized pathways for top international talent to work in the United States. Government should also consider how to incentivize the use of AI models themselves to differentially accelerate reliability and security research (Carlsmith 2025). In the long term, government should redouble its efforts to expand the domestic AI talent pipeline, granting scholarships and research grants to AI undergraduate and graduate students.

### Expanding researcher access

Finally, a lack of research may reflect limited researcher access to advanced AI systems that are either necessary for conducting meaningful research or that are the subject of the research itself. In addition to continuing to use NAIRR, policymakers can democratize access by encouraging industry to waive model access costs for underresourced research organizations. In addition, government can encourage industry to grant external researchers pre-deployment access to models through mechanisms such as regulatory sandboxes or by offering tax incentives to companies that submit their models to independent evaluations.

## Future directions

Refining our understanding of expert priorities necessitates methodological advancements in future elicitation efforts. Building on the limitations identified in this study, subsequent research should aim for:

- **Broader and More Systematic Sampling:** Employing methods to achieve higher response rates and potentially more representative samples of relevant expertise. Incorporating modest incentives, as suggested by Grace (2024), may be beneficial.
- **Refined Survey Instruments:** Improving the clarity and precision of demographic questions (e.g., regarding relevant experience, organizational roles).
- **Exploring Rationales:** Conducting studies (perhaps using qualitative methods or follow-up workshops) to delve deeper into the reasoning behind expert assessments, key considerations, and perceived barriers or enablers for different AI R&S approaches.
- **Inclusive Processes:** Recognizing that technical expertise is only one component, future work should explore methods for incorporating perspectives from diverse stakeholders and the broader public, potentially through participatory workshops or public surveys (see Pasquini, 2025), to ensure alignment with societal values.
- **Mitigating respondent bias:** Address the challenge of respondents' bias toward their own research, either via broader sampling (mentioned above) or by asking respondents to report their area(s) of expertise and including this information in analysis.
- **Real-time Surveying:** The process of designing, administering, and publishing a survey is time-intensive, and results can become quickly outdated. We envision that a more up-to-date version of this survey could be continuously maintained and updated, and plan to explore the feasibility of such a project.[7]

Addressing these methodological challenges will be crucial for developing and implementing a robust portfolio of technical and governance approaches needed to navigate the risks and opportunities of advancing AI.

[7] We also believe it will be important for respondents to add research areas. For example, respondents in this survey suggested 16 additional unique research directions in the qualitative feedback section, suggesting that our initial list did not comprehensively cover the full space of technical AI R&S interventions under active research.

# Conclusion

**Navigating the complex landscape of AI reliability and security research demands clear prioritization**, especially as transformative capabilities develop at a rapid pace. Our survey offers a reference point by **quantifying expert judgments across a wide set of technical research directions**, identifying potential priorities for funders, researchers, and policymakers seeking to mitigate risks from advanced AI.

A clear picture emerged from expert assessments: **immediate priority lies heavily in practical evaluation, monitoring, and forecasting** to anticipate and detect potentially dangerous capabilities before they cause harm. **Multi-agent safety** and **robust applied security measures** also featured prominently. Notably, experts consistently favored concrete, domain-specific interventions over more abstract theoretical frameworks for driving near-term progress. While leading AI companies are active in some top-ranked areas, our findings underscore significant opportunities for additional, targeted support. Furthermore, expert insights highlight the need for sustained, potentially large-scale investment in important but less tractable areas, many of which involve improving security around the development, storage, and access of AI systems.

To maximize the impact of limited resources, **funders and policymakers should leverage both the overall promise scores and the specific bottlenecks identified here**. Strategic deployment of direct funding, investment incentives, researcher support, and field-wide coordination efforts can all contribute to better targeting critical gaps.

We must acknowledge the limitations of this study. First, this study represents a snapshot in a rapidly evolving field. Second, our sample size and response distribution limited our ability to glean insights across all research areas. Future iterations should pursue broader sampling, and explore real-time participatory methods to align research directions with up-to-date priorities.

**We see this work as a template for iterative improvement,** and actively welcome feedback and collaboration to refine our methodology, broaden our reach, and maximize the impact of future iterations. If you have suggestions or wish to contribute, please reach out.

# Acknowledgements

This project benefited significantly from the contributions and expertise of numerous individuals. Assessing the landscape of AI reliability and security is an inherently interdisciplinary and ambitious endeavor, made possible only through generous collaboration. We extend our sincere gratitude to the following individuals whose insights and thoughtful feedback greatly enhanced the quality and rigor of this work: Usman Anwar, Richard Ren, Peter Barnett, Zoe Williams, Ari Holtzman, Oscar Delaney, Micah Carroll, Buck Shlegeris, Daniel Kang, Peter Hase, Daniel Brown, Nouha Dziri, Chirag Agarwal, Niloofar Mireshghallah, Leilani Gilpin, David Krueger, Noemi Dreksler, Willem Sleegers, David Moss, and, of course, all of our anonymous survey respondents. All remaining errors are our own.

# Appendix A: Taxonomy

For this survey, we developed a taxonomy of technical reliability and security research consisting of 105 sub-areas organized into 20 broader categories. See Methods section for details on the design process. We reproduce here the taxonomy in its entirety, along with the descriptions (as seen by survey respondents) and references to example work (which were not shown to respondents).

We believe this taxonomy may serve as a capable basis for ongoing mapping of this research landscape.

## Top-level categories

Respondents first selected one or more of the following categories:

1. Theoretical foundations and provable safety in AI systems
2. Training and fine-tuning methods for alignment and safety
3. Scalable oversight and alignment techniques
4. Understanding in-context learning, reasoning, and scaling behavior
5. Interpretability, explainability, and transparency
6. Robustness
7. Improving the science of AI evaluation
8. Domain-specific AI evaluation design
9. Agentic LLMs and single-agent risks
10. Multi-agent interactions
11. Cooperative AI and mechanism design
12. Fairness
13. Accountability
14. Ethics
15. Choosing and operationalizing values in AI
16. Privacy
17. Cybersecurity for AI models
18. Hardware and infrastructure security for AI
19. Improving general understanding of deep learning
20. Research on safety in non-LLM systems

## Category and sub-area descriptions

Area descriptions were shown to respondents alongside importance and tractability questions. The initial survey did not include "example work" citations for each sub-area, but they have been provided here as additional context for readers.

**1. Theoretical foundations and provable safety in AI systems:** Advancing the theoretical foundations of AI safety by building models and frameworks that ensure provably correct and robust behavior. These efforts span from verifiable architectures and formal verification methods to embedded agency, decision theory, incentive structures aligned with causal reasoning, and control theory.

a. **Building verifiable and robust AI architectures:** Constructing AI systems with architectures that support formal verification and robustness guarantees, such as world models that enable safe and reliable planning, or guaranteed safe AI with Bayesian oracles. This area emphasizes simplicity and transparency to aid in provability. Example work includes (Dalrymple et al. 2024) and (Bengio et al. 2024).
b. **Formal verification of AI systems:** Applying formal methods to verify that AI models and algorithms meet stringent safety, robustness, and performance criteria. This includes proving resilience against adversarial inputs and perturbations, and certifying conformance to specified safety properties under varying conditions. Example work includes (Seshia, Sadigh, and Sastry 2022) and (Henzinger, Lechner, and Žikelić 2021).
c. **Decision theory and rational agency:** Establishing formal decision-making frameworks that ensure rational and safe choices by AI agents, potentially drawing on concepts like causal and evidential decision theory. Example work includes (Tennant, Hailes, and Musolesi 2023).
d. **Embedded agency:** Explores how agents can model and reason about themselves and their environment as interconnected parts of a single system, addressing challenges like self-reference, resource constraints, and the stability of reasoning processes. This includes tackling problems arising from the lack of a clear boundary between the agent and its environment. Example work includes (Demski and Garrabrant 2020).
e. **Causal incentives:** Developing frameworks that formalize how to align agent incentives with safe and desired outcomes by ensuring their causal understanding matches intended objectives. This research provides a formal language for guaranteeing safety, addressing challenges like goal misspecification, and complementing broader efforts in agent foundations and robust system design. Example work includes (Everitt et al. 2021), (Farquhar, Carey, and Everitt 2022), and (Ganguly et al. 2023).
f. **Control theory applications in AI safety:** Leveraging principles from control theory to ensure stability, robustness, and safety for AI-driven systems interacting with dynamic physical environments. This includes designing controllers and feedback mechanisms to maintain system integrity, prevent runaway behaviors, and achieve desired performance criteria under uncertainty. Example work includes (Xiao et al. 2024) and (D. D. Fan et al. 2020).

**2. Training and fine-tuning methods for alignment and safety:** Developing reliable training and fine-tuning strategies for AI models to ensure that their outputs remain safe, interpretable, and aligned with intended goals. This involves understanding how fine-tuning affects model behavior, employing adversarial training for robust alignment, carefully adjusting pre-training processes, and improving data quality and auditing methods.

a. **Understanding how fine-tuning changes a pretrained model:** Investigating how fine-tuning alters a model's internal representations and behaviors to better predict, and ultimately control, downstream safety outcomes. Example work includes (B. Y. Lin et al. 2023), (Jain et al. 2024), and (Clymer et al. 2023).
b. **Develop output-based adversarial training techniques for more robust alignment:** Developing training procedures, such as adversarial training focused on internal model representations, or 'process supervision', that directly optimize against adversarial examples and undesirable outputs, making models more resistant to manipulations that could lead to unsafe behaviors. Example work includes (Miyato et al. 2019), (Lightman et al. 2023), and (Casper et al. 2024).
c. **Scalable techniques for targeted modifications of LLM behavior (including unlearning):** Creating scalable methods for precisely adjusting model outputs, such as removing unwanted content or refining responses to adhere to alignment constraints without broadly degrading performance. This may also include removal of *unknown* or latent undesirable capabilities that emerge in large models. Example work includes (Hubinger et al. 2024), (Y. Yuan et al. 2024), (Belrose et al. 2025), and (Jang et al. 2023).
d. **Retrieval-augmented pre-training:** Incorporating retrieval mechanisms during pre-training to better ground models in verified information. Example work includes (Guu et al. 2020).
e. **Pretraining alterations to improve interpretability:** Altering pre-training protocols to produce models with clearer internal representations and decision-making pathways, allowing for more effective downstream analysis and intervention. Example work includes (Ismail, Bravo, and Feizi 2021) and (Golechha, Cope, and Schoots 2024).
f. **Limiting models' ability to perform harmful tasks:** Introducing mechanisms during pre-training that proactively limit a model's potential to learn or perform harmful tasks, constraining the model's capability space to safer domains before downstream fine-tuning. Example work includes (Henderson et al. 2023) and (Zhou et al. 2023).
g. **Scalable data auditing, filtering, and Pretraining with Human Feedback (PHF):** Developing tools for large-scale data auditing, filtering, training-data attribution, and incorporating human feedback at the pre-training stage. Example work includes (Fernando et al. 2023; Agarwal, D'souza, and Hooker 2022; Grosse et al. 2023; and Korbak et al. 2023).

**3. Scalable oversight and alignment techniques:** Developing approaches to guide and align increasingly complex AI systems even in tasks where direct oversight is challenging, such as by the use of AI feedback, debate, iterative training processes, and enhanced elicitation methods.

a. **Reinforcement Learning from AI Feedback (RLAIF):** Using feedback generated by AI systems to guide reinforcement learning, effectively scaling the oversight process beyond purely human-labeled data. Example work includes (H. Lee et al. 2023).
b. **Debate:** Encouraging multiple models (or model instances) to discuss and critique each other's reasoning, with human overseers judging the best arguments. Example work includes (Brown-Cohen, Irving, and Piliouras 2023) and (Irving, Christiano, and Amodei 2018).
c. **Iterated Distillation and Amplification (IDA):** An alignment approach where increasingly capable AI systems are trained by recursively using weaker AIs to teach and amplify smarter successors. To address the limitations of human-defined feedback and reward functions, IDA decomposes complex tasks—using AI assistance—into simpler subtasks with accessible human or algorithmic evaluation signals, enabling scalable alignment and improved performance over time. Example work includes (Christiano, Shlegeris, and Amodei 2018).
d. **Better elicitation mechanisms from humans:** Improving methods to extract more reflective, aspirational, and consistent human preferences, to provide data to guide AI systems along these preferences and update in accordance with changes in values over time. Example work includes (Klingefjord, Lowe, and Edelman 2024) and (Oliveira et al. 2023).
e. **Recursive reward modeling:** Breaking down complex tasks into simpler subtasks for which reward signals can be more easily specified, then "building up" to oversee more complex behaviors. Example work includes (Leike et al. 2018) and (Jeff Wu et al. 2021).

**4. Understanding in-context learning, reasoning, and scaling behavior:** Methods to gain a comprehensive understanding of how large language models learn, reason, and scale, such as by examining in-context learning (ICL) mechanisms, the influence of data and design on behavior, the theoretical foundations of scaling, the emergence of advanced capabilities, and the nature of reasoning.

a. **Mechanistic understanding of In-Context Learning:** Investigating the internal processes by which transformers perform ICL, including whether these processes resemble emergent optimization behavior, advanced pattern-matching, or other structural mechanisms. This research may include scenario-based analyses to identify the circuits critical for ICL under artificial constraints. Example work includes (Xie et al. 2021), (L. Lin, Bai, and Mei 2024), and (Olsson et al. 2022).
b. **Influences on ICL behavior and performance:** Examining how the tasks, instructions, pre-training data distribution, and design choices (e.g., instruction tuning, model size, training duration) shape the range and reliability of behaviors that can be specified in-context. Example work includes (Y. Wang et al. 2023) and (Hahn and Goyal 2023).
c. **Theoretical and representational aspects of scaling:** Clarifying when and how scaling drives improvements, such as by building a more robust theoretical framework to describe scaling laws, or analyzing how increasing model size and training data influence learned representations. Example work includes (Viering and Loog 2023) and (Vyas et al. 2023).

d. **Emergence and task-specific scaling patterns:** Formalizing and forecasting the emergence of new capabilities as models scale, investigating whether scaling alone can produce certain capabilities, and designing methods for discovering task-specific scaling laws. Example work includes (McKenzie et al. 2024), (Ganguli et al. 2022), and (Caballero et al. 2023).
e. **Impact of scaling and training on reasoning capabilities:** Determining whether and how increases in model size and training complexity enhance reasoning abilities, and identifying which aspects of training conditions and data sources facilitate the acquisition of reasoning skills. Example work includes (Wei et al. 2022), (Saparov et al. 2023), and (Magister et al. 2023).
f. **Mechanistic understanding and limits of LLM reasoning:** Examining the underlying mechanisms of reasoning in LLMs, exploring non-deductive reasoning capabilities of LLMs (e.g., causal or social reasoning). Example work includes (Hou et al. 2023) and (Gandhi, Sadigh, and Goodman 2023).
g. **Limits of Transformers:** Defining the computational limits of transformers in supporting sophisticated reasoning. Example work includes (Merrill and Sabharwal 2023) and (Strobl 2023).

**5. Interpretability, explainability, and transparency:** Ensuring that AI systems are understandable, trustworthy, and transparent. This involves developing tools and methods to interpret model internals, refining the reliability and scalability of interpretability techniques, exploring ways to elicit and explain model reasoning, and improving the transparency of complex models.

a. **Interpretability foundations:** Focuses on theoretical and experimental studies investigating how models represent and encode concepts, emphasizing structural and abstraction-level insights, including by distinguishing linear from non-linear encodings, understanding polysemanticity and superposition, examining concept mismatches between models and humans, and discovering more accurate abstractions for interpretability. Example work includes (Bilodeau et al. 2024), (Mahinpei et al. 2021), and (Elhage et al. 2022).
b. **Validating and applying interpretability methods:** Developing rigorous criteria and benchmarks for evaluating the reliability of interpretability methods, and understanding whether these methods maintain their validity when applied to actively modify model behavior. Example work includes (K. Wang et al. 2022) and (Schwettmann et al. 2023).
c. **Feature and circuit analysis:** Creating scalable approaches for feature interpretation, circuit discovery, and feature steering (or top-down/control vectors). Example work includes (Nanda et al. 2023) and (Conmy et al. 2023).
d. **Eliciting Latent Knowledge (ELK):** Developing methods to reveal hidden knowledge embedded within models, enabling researchers to identify what models implicitly “know” about the world and how this knowledge influences predictions. Example work includes (Christiano 2023).

e. **Developmental interpretability:** Investigating how AI models' internal representations and behaviors evolve throughout their training process to understand the developmental stages and mechanisms by which complex capabilities emerge. This research aims to uncover the progressive changes in model structure and function, facilitating better alignment and safety assurances. Example work includes (G. Wang et al. 2024) and (Hoogland et al. 2025).

f. **Transparency:** Research here aims to open up the black box of AI systems by uncovering how data, architecture, and training processes shape model outputs. Example research focuses on advanced documentation frameworks, auditing tools to surface biases or vulnerabilities, and reporting protocols to effectively explain outputs and communicate uncertainty. Example work includes (Felzmann et al. 2020), (Liao and Vaughan 2024), and (Bommasani et al. 2023).

g. **Explainability:** Methods to understand why a model generates specific outputs. Technical approaches include developing post-hoc or embedded explanation methods, measuring and improving explanation fidelity, and crafting user-focused interfaces that clarify causal or logical relationships. Example work includes (Madsen, Chandar, and Reddy 2024) and (Y. Wu et al. 2022).

**6. Robustness:** Ensuring that AI systems remain reliable and secure in the face of adversarial manipulation, misaligned inputs, and uncertain conditions, such as by protecting against prompt-based exploits, poisoning attacks, and adversarial perturbations, and introducing control mechanisms and uncertainty quantification methods to maintain resilient system behavior at scale.

a. **Defending against jailbreaks and prompt injections:** Improving state-of-the-art methods for discovering, evaluating, and defending against prompt injection and "jailbreaking" attacks. Research also focuses on structural defenses, such as detection, filtering, and paraphrasing of prompts, as well as addressing vulnerabilities stemming from a lack of robust privilege levels (e.g., system prompt vs. user instruction) in LLM inputs. Example work includes (Zou et al. 2023), (Bailey et al. 2024), and (Han et al. 2024).

b. **Defending against poisoning and backdoors:** Understanding how LLMs can be compromised through data poisoning at various training stages, examining the effect of model scale on vulnerability, testing out-of-context reasoning under poisoning, and exploring attacks via additional modalities and encodings. This area also includes detecting and removing backdoors (i.e., Trojan detection) to ensure that covertly embedded harmful behaviors are mitigated. Example work includes (Carlini et al. 2024), (Qi, Xie, et al. 2023), (Wan et al. 2023), and (Hubinger et al. 2024).

c. **Adversarial robustness to perturbations:** This area investigates how models can be made more resilient to carefully crafted adversarial perturbations designed to degrade performance or reveal vulnerabilities. Research involves identifying methods for bolstering model robustness under challenging conditions, including adversarial training and certified defenses. Example work includes (Nandi et al. 2023) and (Addepalli et al. 2022).

d. **Control mechanisms for untrusted models:** Designing and evaluating protocols to control outputs from untrusted models. This includes methods for monitoring backdoored outputs, integrating control measures with traditional insider risk management strategies, building safety cases for control tools, and employing white-box techniques (e.g., linear probes) for continuous oversight. Example work includes (Greenblatt et al. 2024).
e. **Uncertainty quantification:** Quantifying uncertainty in model predictions. Techniques include ensemble methods, conformal predictions, and Bayesian approaches to estimate and calibrate model confidence. Example work includes (Y. Hu et al. 2022) and (Shaker and Hüllermeier 2021).

**7. Improving the science of AI evaluation:** Ensuring that AI systems can be accurately assessed and understood. This includes theoretical work in capability and safety evaluation, as well as improving the reliability and fairness of evaluation processes.

a. **Theoretical foundations for evaluation:** Research includes: creating rigorous frameworks for predicting capabilities (as opposed to relying solely on benchmarks) understanding generality and generalization in LLMs, and developing theory-grounded taxonomies of model capabilities. Example work includes (Burnell et al. 2023) and (Schaeffer et al. 2025).
b. **Evaluation methodology and metrics:** Research focuses on designing holistic, theory-grounded metrics (e.g., focused on more than just harmlessness), accounting for scaffolding in evaluations, and characterizing safety–performance trade-offs. Example work includes (Liang et al. 2023) and (Kiela et al. 2021).
c. **Studying misalignment through simplified model organisms:** Developing and studying simplified AI models—"model organisms"—to probe potential misalignment, gain insights into failure modes, and refine evaluation strategies without the complexity of full-scale systems. Example work includes (Hubinger et al. 2024).
d. **Improving evaluation robustness:** Methods here aim to stabilize evaluations against sensitivity to prompts, detect and address contaminated data, ensure that evaluations remain meaningful even if models are fine-tuned in a targeted manner for certain tasks, and mitigate bias in AI evaluations (including biases in crowdsourced human evaluations). Example work includes (Sclar et al. 2024), (Golchin and Surdeanu 2024), (M. Wu and Aji 2023), (Yong, Menghini, and Bach 2024) and (Z. Wu et al. 2024).
e. **Detecting and addressing previously unmeasured or latent capabilities:** Developing strategies to uncover latent harmful abilities within AI models and prevent models from exhibiting undesirable behaviors such as "sandbagging" or deceptively underperforming during evaluations. Example work includes (Ward et al. 2023) and (van der Weij et al. 2025).

**8. Domain-specific AI evaluation design:** Developing specialized evaluation tools to assess AI models' capabilities and safety in critical areas such as automated AI research and development, cybersecurity, chemical/biological/radiological/nuclear (CBRN) scenarios, and manipulative behaviors like deception and persuasion.

a. **Automated AI R&D evaluations:** Designing evaluations to assess a model's capacity to generate research ideas, propose improvements to algorithms, or autonomously advance AI capabilities. Example work includes (Wijk et al. 2024) and (Owen 2024).
b. **Cyber evaluations:** Designing evaluations to assess a model's ability to understand, exploit, or defend against cybersecurity threats and vulnerabilities. Example work includes (AI Security Institute 2024).
c. **CBRN (Chemical, Biological, Radiological, and Nuclear) evaluations:** Designing evaluations to assess a model's understanding of hazardous CBRN materials and scenarios, ensuring it cannot be easily leveraged to facilitate harmful acts involving these agents. Example work includes (AI Security Institute 2024) and (Laurent et al. 2024).
d. **Evaluating deception, scheming, situational awareness, and persuasion:** Designing evaluations to assess how well models can deceive, strategize, maintain situational awareness, or influence human decision-making. Example work includes (Scheurer, Balesni, and Hobbhahn 2024).

**9. Agentic LLMs and single-agent risks:** Developing a deeper understanding of agentic behavior in LLM-based systems. This work clarifies how LLM-agents learn over time, respond to underspecified goals, and engage with their environments.

a. **Lifelong learning and goal-directedness in LLM agents:** Investigating how agentic LLMs evolve through ongoing learning, and potentially exhibit undesirable behaviors due to goal-directedness. Example work includes (Qi, Zeng, et al. 2023) and (Perez et al. 2023).
b. **Robustness to underspecification:** Enhancing methods to ensure LLM-agents remain aligned despite vague or shifting objectives. Example work includes (Mu et al. 2023) and (Turner, Hadfield-Menell, and Tadepalli 2020).
c. **Oversight and monitoring of LLM-agents:** Building automated oversight and monitoring tools to track LLM-agent actions. Example work includes (Naihin et al. 2023) and (A. Chan et al. 2024).
d. **Evaluating tool affordances for LLM-agents:** Evaluating the safety of providing LLM-agents with tools and affordances, and determining whether robust safety assurances are possible for given affordances. Example work includes (Ruan et al. 2024).

**10. Multi-agent interactions:** Research focusing on ensuring safe multi-agent interactions, such as by detecting and preventing malicious collective behaviors, studying how transparency can affect agent interactions, and developing evaluations for agent behavior and interaction.

a. **Safety and emergent functionality in multi-agent interactions:** Understanding how individual agent dispositions and capabilities scale into complex multi-agent dynamics, evaluating emergent functionalities (e.g., coordinated strategies), enhancing robustness of LLM agents to correlated failures stemming from foundationality, and applying insights from multi-agent RL research to LLM-based systems. Example work includes (Critch, Dennis, and Russell 2022), (Juneja et al. 2024), and (Yocum et al. 2023).

b. **Detecting and preventing collusion and emergent collective behavior:** Developing detection techniques (e.g., information-theoretic or interpretability-based) for collusion between AI agents, benchmarking and evaluating collusive tendencies, designing mitigation strategies such as oversight regimes, communication restrictions, and methods for steering agents, understanding conditions (e.g., agent similarity, communication channels, environment structure) that facilitate collusion, and understanding why and how general "super-agents" might develop from many narrow agents. Example work includes (Motwani et al. 2025), (Calvano et al. 2020), and (Ha and Tang 2022).
c. **Multi-agent security:** Assessing unique security risks that arise in multi-agent ecosystems, designing defenses (e.g., secure communication protocols, improved network architectures, information security), studying how multiple systems can circumvent safeguards, evaluating robustness of cooperation to adversarial attacks (e.g., if a small number of malicious agents can destabilize larger groups), evaluating how well agents can adversarially attack each other, and studying the impact of AI agent's training dynamics on data generated by each other with respect to shared vulnerabilities/correlated failure modes. Example work includes (D. Zhang et al. 2021), (D. Lee and Tiwari 2024), and (Jones, Dragan, and Steinhardt 2024).
d. **Network effects and destabilizing dynamics in agent ecosystems:** Understanding which network structures and interaction patterns lead to robust or fragile systems, monitoring and controlling dynamics and co-adaptation of networks of advanced agents, and identifying important security concerns in existing and future multi-agent application areas (e.g., finance, energy grids) and applying lessons from those areas to manage destabilizing forces. Example work includes (Leonardos and Piliouras 2020), (Bloembergen et al. 2015), and (Sanders, Farmer, and Galla 2018).
e. **Transparency, information asymmetries, and communication protocols:** Studying how agent transparency (e.g., code access) or predictability of agents can influence cooperation or defection, scaling Bayesian persuasion and information design to complex multi-agent settings, developing secure information transmission methods between AI agents to promote cooperation,, examining how agent similarity and evidential reasoning about others affect ability and propensity to cooperate, and developing efficient algorithms for zero- or few-shot coordination in high-stakes scenarios. Example work includes (Conitzer and Oesterheld 2023), (Kamenica 2019), and (H. Hu et al. 2021).
f. **Multi-agent metrics and evaluations:** Distinguishing and measuring cooperative dispositions, understanding agents' robustness against coercion or exploitation, quantifying traits like altruism or spite, assessing the impact of capability asymmetries between agents, examining how training processes and data sources influence cooperation, and developing dangerous capability evaluations for multi-agent systems. Example work includes (Abdelnabi et al. 2023), (Agapiou et al. 2023), and (Mukobi et al. 2023).

**11. Cooperative AI and mechanism design:** Fostering beneficial multi-agent ecosystems through research on human-AI interaction, mechanism design, communication protocols, peer incentivization, and automated mechanism design.

a. **Human-AI interaction and collaboration:** Designing AI systems that can understand and predict human actions and preferences; creating interfaces and protocols for effective human-AI teamwork; understanding how interactive AI may change human decision making. Example work includes (Leitão et al. 2022) and (Alon-Barkat and Busuioc 2023).
b. **Mechanism design and multi-agent communication:** Focuses on foundational concepts like social choice theory, incentive alignment, and emergent communication protocols in multi-agent systems to ensure cooperation and fair outcomes. Example work includes (Guresti, Vanlioglu, and Ure 2023) and (Brandt et al. 2016).
c. **Peer incentivisation and automated mechanism design:** Focuses on practical and scalable applications of mechanism design, including methods for incentivizing cooperation among agents, designing secure and scalable inter-agent contracting and norm enforcement mechanisms, and structured opponent-shaping strategies in complex environments. Example work includes (J. Yang et al. 2020), (Eccles et al. 2019), and (Foerster et al. 2018).

**12. Fairness:** Research focusing on developing equitable AI systems, including detecting and mitigating bias, ensuring fair representation across diverse groups, addressing fairness in dynamic or constrained data scenarios, and reconciling conflicting fairness definitions to align interventions with societal values.

a. **Fairness under dynamic and constrained data scenarios:** Ensuring that fairness interventions remain effective under continual learning, adaptive deployment, or evolving operational contexts. Example work includes (Zhao et al. 2022) and (Zhao et al. 2023).
b. **Fair representation and participation in AI systems:** Promoting fair representation and generalization across different subpopulations, and ensuring inclusive participation in the development and governance of AI systems. Example work includes (Birhane et al. 2022) and (Kirk et al. 2024).
c. **Bias detection, quantification, and mitigation techniques:** Developing systematic methods to detect, measure, and reduce bias in model outputs, ranging from pre-processing adjustments to post-hoc corrections. This may also include causal methods for fairness, such as causal modeling techniques to distinguish between genuine causal relationships and spurious correlations in observed disparities, enabling fairness interventions that address underlying structural causes. Example work includes (Plecko and Bareinboim 2022), (Binkytė-Sadauskienė et al. 2022), (Hort et al. 2023) and (Mehrabi et al. 2021).
d. **Fairness in multilingual, cross-cultural, and multimodal contexts:** Addressing fairness challenges that arise when models operate across different languages, cultures, and data modalities. Example work includes (Adewumi et al. 2024) and (Tao et al. 2024).

e. **Intersectional fairness and complex group structures:** Addressing compounded biases that arise when protected attributes overlap, such as race and gender, to ensure fairness approaches capture nuanced harms across intersectional groups. This research develops computational methods and evaluation frameworks to avoid oversimplifying population categories and to identify disparities affecting complex group structures. Example work includes (Gohar and Cheng 2023) and (W. Guo and Caliskan 2021).
f. **Reconciling multiple fairness definitions and normative trade-offs:** Comparing and combining conflicting formal definitions of fairness to address the normative trade-offs they entail and align fairness interventions with societal values. This research clarifies the theoretical and practical implications of fairness definitions, helping practitioners navigate complex policy and ethical considerations. Example work includes (Buijsman 2023) and (Bateni, Chan, and Eitel-Porter 2022).

**13. Accountability:** Research focusing on ensuring AI systems are transparent, reliable, and compliant, including developing auditing tools, attributing AI outputs to specific models, mitigating risks of power concentration in AI development and deployment, and automating regulatory compliance.

a. **Auditing mechanisms:** Developing automated post-hoc auditing tools, using privacy-enhancing technology to facilitate secure audit access to sensitive data, building auditability into systems by design, and establishing continuous accountability pipelines that monitor, log, and assess model behaviors over time to support transparent and verifiable assessments of model behaviors. Example work includes (Marone and Durme 2023), (Ilyas et al. 2022) and (Bluemke et al. 2023).
b. **Methods for detecting and attributing LLM outputs:** Developing techniques—such as watermarking or model fingerprinting—to identify and attribute content to its source model provides a foundation for accountability, reduces misinformation risks, and clarifies responsibility. Example work includes (Kirchenbauer et al. 2024) and (J. Xu et al. 2024).
c. **Regulatory compliance automation:** Automating processes to ensure that models conform to legal standards, industry guidelines, and ethical principles helps organizations proactively meet accountability requirements while reducing manual oversight burdens. Example work includes (Cappelli and Di Marzo Serugendo 2025) and (Sojasi 2024).
d. **Methods for mitigating power concentration in AI:** Investigating mechanisms to prevent the centralization of AI capabilities and influence—such as decentralized governance, open-source contributions, and equitable resource allocation. Example work includes (Y. Liu et al. 2024) and (Montes and Goertzel 2019).

**14. Ethics:** Work on AI ethics includes developing methods for integrating ethical considerations into training, evaluation, and decision-making processes, as well as techniques for mitigating harmful outputs and ensuring cultural and long-term ethical consistency.

a. **Ethics-aware training and fine-tuning:** Research on learning from imperfect ethical datasets, applying ethics-aware data curation methods, and incorporating collective ethical principles into model design. Example work includes (Hendrycks et al. 2023) and (Lourie, Bras, and Choi 2021).
b. **Ethical decision-making frameworks:** Developing formal risk-aware, algorithmic harms assessment, and domain-specific ethical decision-making frameworks tailored for large language models and related AI systems. Example work includes (Tennant, Hailes, and Musolesi 2023) and (Ji et al. 2024).
c. **Mitigating harmful outputs:** Approaches include refining models to reduce the production of dangerous, misleading, or otherwise harmful outputs, employing filtering, red-teaming, and reinforcement learning from human feedback. Example work includes (Dai et al. 2023) and (Ngo et al. 2021).
d. **Cultural sensitivity and contextual awareness:** Techniques aim to adapt models to diverse cultural contexts and subtle social norms, ensuring that outputs remain appropriate, respectful, and aligned with local values. Example work includes (Li et al. 2024) and (A. J. Chan et al. 2024).
e. **Long-term ethical consistency:** Research explores methods for maintaining stable, ethically coherent model behavior over extended periods, including approaches to prevent drift and to preserve core ethical principles despite shifting inputs. Example work includes (McAllister et al. 2018), (Christiano et al. 2023), and (Everitt et al. 2021).

**15. Choosing and operationalizing values in AI:** This area focuses on developing principled methods to identify, justify, and implement value systems within AI models, reconciling diverse ethical priorities, managing conflicts, and creating robust evaluations to ensure models embody chosen values.

a. **Justify value choices for alignment**: Research includes formulating principled criteria and philosophical foundations that guide why certain values should be encoded into AI systems. Example work includes (Gabriel 2020) and (Zhi-Xuan et al. 2024).
b. **Manage conflicts between different values:** Efforts here explore approaches like multi-objective optimization or deliberation frameworks to resolve cases where multiple values clash. Example work includes (Qu et al. 2021) and (Kirk et al. 2024).
c. **Develop more robust evaluations for which values an LLM encodes:** Researchers design metrics and tests to identify and measure the values present in a model's behavior, outputs, and decision-making processes. Example work includes (Scherrer et al. 2023) and (Arora, Kaffee, and Augenstein 2023).
d. **Pluralistic value alignment:** Strategies focus on simultaneously accommodating multiple, possibly diverse value systems, enabling AI to adapt to different stakeholders or cultural contexts. Example work includes (Sorensen et al. 2024).
e. **Foundational research on operationalizing values in LLMs:** This includes theoretical and empirical studies on how to incorporate values directly into training procedures, fine-tuning protocols, and model architectures. Example work includes (Solaiman and Dennison 2021) and (C. Xu et al. 2023).

**16. Privacy:** This area focuses on identifying and mitigating privacy risks arising from new capabilities and deployment scenarios for LLMs, developing robust conceptual frameworks for privacy definitions, and leveraging AI tools to preserve and enhance privacy in various application domains.

a. **Identifying emergent privacy risks in new paradigms:** Examining novel attack vectors (e.g., inference time risks) in new paradigms (e.g., retrieval-augmented generation, agent-based interactions, plugin ecosystems) to uncover how these integrations may lead to unexpected disclosures. Example work includes (Jegorova et al. 2022), (Mireshghallah et al. 2024), and (Zeng et al. 2024).
b. **Research on inferring sensitive information from accumulation of innocuous data:** Studying how seemingly harmless data points can be combined to reconstruct sensitive information, enabling adversaries to "weaponize" aggregate inferences against individuals. Example work includes (Kröger 2019) and (Staab et al. 2024).
c. **Privacy challenges in complex data scenarios:** Exploring how complex data scenarios, such as cross-lingual and cross-modal transformations (e.g., images, audio, code snippets) can reveal protected content, examining what can be extracted from data presented in alternative formats. This also includes work on context-specific privacy norms, ensuring that privacy measures adapt to different cultural, social, and situational factors rather than relying on one-size-fits-all policies. Example work includes (Nissenbaum 2010), (W. Fan et al. 2024), (Huang et al. 2024), and (Choi, Min, and Choo 2024).
d. **Privacy modeling frameworks:** Developing more precise models of privacy that align with user expectations—potentially informed by human-computer interaction (HCI) research—and grounding these definitions in implementable policies. This also includes formalizing methodologies to characterize and prioritize the worst-case privacy outcomes, moving beyond ad hoc assessments and towards systematic threat modeling frameworks. Example work includes (Mollaeefar, Bissoli, and Ranise 2024), (Villegas-Ch and García-Ortiz 2023), (Brown et al. 2022) and (Mireshghallah et al. 2024).
e. **Data encryption tools for model inputs and outputs:** Techniques for encrypting inputs, outputs, and intermediate representations during runtime to ensure confidentiality and prevent unauthorized access to sensitive queries or responses. Example work includes (Mishra, Li, and Deo 2023) and (S. Lin et al. 2025).

**17. Cybersecurity for AI models:** Focuses on protecting model parameters, interfaces, training techniques, and outputs from unauthorized access, extraction, or misuse using cryptographic, architectural, and procedural safeguards. This includes ensuring secure weight storage, hardened access control, oracle protection measures, protecting algorithmic insights, preventing self-exfiltration, and robust data integrity.

a. **Weight security and key management:** Research focuses on cryptographic techniques for encrypting and securely storing model weights at rest and in transit, hardware-based protections (e.g., trusted execution environments) that ensure the model's parameters cannot be extracted even with physical access, developing systems for isolating weights behind tightly controlled interfaces, and implementing formal verification of key generation, storage, and rotation protocols. Example work includes (Nevo et al. 2024b) and (Aarne, Fist, and Withers 2024).

b. **Access control and interface hardening:** Approaches include creating minimal, verifiable interfaces for weight access, deploying multiparty authorization and cryptographic attestation protocols to guard against model extraction, novel authentication and authorization schemes integrating Zero Trust principles at a granular level (beyond standard Identity & Access Management [IAM] tools), and implementing AI firewalls with strict input-output validation. Example work includes (Nevo et al. 2024b).
c. **Model robustness and oracle protection:** Techniques prevent model extraction through inference-only attacks, detect and filter adversarial inputs designed to reconstruct the model or degrade its integrity, and employ adversarial training and advanced input/output “reconstruction” methods to limit the risk that internal representations are inferred from model queries. Example work includes (Carlini and Wagner 2017) and (Carlini et al. 2019).
d. **Preventing model self-exfiltration:** Methods to ensure that models cannot covertly leak sensitive information about their internal parameters or training data, or copy their own weights to external devices and networks, such as output restrictions, sanitization techniques, or fine-grained monitoring of responses. Example work includes (Clymer, Wijk, and Barnes 2024), (Kinniment et al. 2024), and (Greenblatt et al. 2024).
e. **Detecting modified models or poisoned data:** Developing methods to detect models that have been maliciously modified or training data which has been poisoned. Example work includes (Sikka et al. 2023) and (di Gaspari, Hitaj, and Mancini 2024).
f. **Quantifying cyber threats from advanced capabilities:** Threat modeling and evaluation of cyber threats from advanced AI models, whether via autonomy or providing human uplift for more "traditional" cyber capabilities. Example work includes (Gupta et al. 2023) and (AI Security Institute 2024).

**18. Hardware and infrastructure security for AI:** Ensuring the security of AI systems at the hardware and infrastructure level involves protecting model weights, securing deployment environments, maintaining supply chain integrity, and implementing robust monitoring and threat detection mechanisms. Methods include the use of confidential computing, rigorous access controls, specialized hardware protections, and continuous security oversight.

a. **Confidential computing and environment isolation:** Using trusted execution environments (such as secure enclaves) to ensure that model weights and computations remain confidential and tamper-proof during large-scale AI inference and training. This also involves reducing the attack surface through sandboxed, code-minimal deployments, specialized hardware/firmware stacks, and maintaining verifiable runtime integrity checks. Example work includes (Dhar et al. 2024) and (Vaswani et al. 2022).
b. **Supply chain integrity and secure development:** Ensuring end-to-end verification of hardware and software supply chains through source-verified firmware, SLSA compliance, and secure software development lifecycles tailored for ML-specific infrastructure. This also includes developing automated tooling to continuously verify the provenance and integrity of model components, dependencies, and third-party code used in training and inference pipelines. Example work includes (Hepworth et al. 2024) and (Mora-Cantallops et al. 2021).

c. **Continuous monitoring, advanced threat detection, and incident response:** Developing ML-driven anomaly detection and logging systems capable of flagging and responding to subtle infiltration attempts or insider threats in real-time. This also includes red-teaming and automated penetration testing frameworks specialized for AI systems, including simulations of zero-day attacks and insider compromises. Example work includes (Cai et al. 2024).
d. **Hardware-integrated monitoring and verification:** Integrating monitoring capabilities directly into hardware, such as secure counters and tamper-evident seals, along with deploying specialized firmware that can detect and respond to attempts at parameter theft or physical attacks. This also includes verification tools, such as hardware-level logging and secured audit trails that remain verifiable under sophisticated tampering attempts for rapid, evidence-based incident response. Example work includes (Petrie, Aarne, and Ammann 2024).
e. **Specialized chips to compute encrypted data:** Designing and deploying hardware accelerators optimized for computations on encrypted data, such as homomorphic encryption schemes, to facilitate efficient encrypted training and inference without exposing plaintext model parameters or sensitive input data outside the protected hardware boundary. Example work includes (Samardzic et al. 2022) and (Aikata et al. 2023).
f. **Tamper-evidence and tamper-proofing:** Implementing tamper-resistant enclosures, seals, and other tamper-evident mechanisms to ensure that any unauthorized physical access or modification attempts are detectable. Such measures help maintain the integrity of hardware components and prevent adversaries from compromising the system at a physical level. Example work includes (Thomaz et al. 2023) and (Vidaković and Vinko 2023).
g. **Datacenter security:** Relevant research focuses on designing and deploying resilient hardware- and software-based defenses to prevent model theft and sabotage. This includes methods like encrypted computation, secure enclaves, continuous anomaly detection, zero-trust architectures, and rigorous supply chain verification to protect against both external intrusions and insider threats. Example work includes (Nevo et al. 2024a).

**19. Improving general understanding of deep learning:** This area focuses on developing rigorous explanations for why deep neural networks learn effectively, uncovering the principles behind generalization, understanding optimization behavior, and analyzing how implicit biases and overparameterization influence performance and safety.

a. **Theoretical foundations of deep learning:** Constructing mathematical models to explain generalization in deep neural networks despite overparameterization, and studying the influence of network architecture on learning properties. Example work includes (Allen-Zhu, Li, and Liang 2019) and (Nagarajan 2021).

b. **Optimization and loss landscape analysis:** Studying the geometry of loss functions and how optimization algorithms navigate them, and examining phenomena such as flat versus sharp minima, and the connection between these properties and robust generalization. Example work includes (Stutz, Hein, and Schiele 2021) and (C. Liu, Zhu, and Belkin 2022).
c. **Implicit bias of optimization algorithms:** Analyzing how algorithms like stochastic gradient descent and related methods influence learned models, and exploring how implicit regularization affects model performance and safety. Example work includes (Jingfeng Wu et al. 2021) and (Chou, Rauhut, and Ward 2024).
d. **Double descent and overparameterization:** Investigating the double descent risk curve and its implications for model capacity, and how overparameterization can lead to improved generalization. Example work includes (Luzi, Dar, and Baraniuk 2024) and (Dar, Luzi, and Baraniuk 2023).

**20. Research on safety in non-LLM systems:** Exploring safety challenges in non-LLM systems, such as robotics and embodied AI, vision and perception systems, and alternative paradigms for developing artificial intelligence (*e.g.*, whole-brain emulation).

a. **Safe reinforcement learning for non-LLM systems:** Developing RL algorithms that prioritize safety during exploration and exploitation, with applications in non-LLM systems such as robotics and embedded AI. This includes incorporating safety constraints and risk-sensitive objectives into learning processes. Example work includes (Brunke et al. 2021) and (F. Yang et al. 2024).
b. **Robotics and embodied AI safety:** Designing robust control systems, fail-safe mechanisms, and dependable sensors for physical systems, such as autonomous vehicles, drones, and household robots, to ensure safe human-robot interaction and accident prevention. Example work includes (W. Zhang et al. 2024).
c. **Adversarial robustness in vision and perception systems:** Studying how malicious inputs can deceive image recognition or sensor-based models, and creating defenses—such as adversarial training, certifiable robustness methods, and detection schemes—to maintain reliable perception. Example work includes (Tu et al. 2022) and (Shao et al. 2022).
d. **Whole-brain emulation:** Exploring the theoretical challenges of accurately replicating a human brain's functionality and ensuring that such emulations—if ever realized—adhere to rigorous safety and ethical standards, avoiding unintended cognitive hazards or harmful behavioral patterns. Example work includes (Duettman and Sandberg 2023) and (Mandelbaum 2022).

# Appendix B: Survey Instructions

This appendix reproduces the information seen by respondents as they encountered the survey.

## Introductory page

Welcome to the AI Assurance and Reliability Research Priorities Survey

**Purpose**
This survey aims to identify high-priority AI assurance and reliability research areas to guide funders—ranging from philanthropic organizations to government agencies—toward impactful funding decisions.

**Your Role**
As an expert in AI/ML or related fields, your input can help highlight which technical challenges, if tackled, would most reduce the risk of severe harm from AI systems (defined as >100 lives lost or >$10 billion in economic impact).

### How the Survey Works

**Select areas of interest (one area ≈ 10 minutes)**
First, choose one or more areas of interest that align with your expertise. **You can choose how much time to invest by selecting more or fewer areas**; more input helps, but is not required. Even the minimum response of choosing one area is extremely valuable, and we thank you for it.

**Rating and feedback**
Then, for the selected area(s), rate the importance and tractability of a handful of sub-areas, and provide brief qualitative feedback.

**Survey Design Details**
There are 20 high-level research areas to choose from, and 3-7 sub-areas per research area.

Items are randomized to minimize order effects.

The survey is anonymous by default, but you may provide your name if you'd like to be contacted for follow-up.

**Outcome**
Your insights will help inform future funding priorities and research agendas as we engage with a broad range of funders and research organizations. We plan to publish aggregated and anonymized results so that researchers, grantmakers, and other interested parties can draw on these insights when shaping their strategies.

**Note on Methodology**

This survey was developed by reviewing recent literature (e.g., Anwar et al., 2024) and consulting with a range of experts. We aimed to cover key technical challenges related to the reliability and security of advanced AI systems (primarily LLMs), but this selection is neither exhaustive nor perfect. While we strived for breadth and relevance, some omissions or biases may remain. We welcome your input to refine our approach for future advocacy for funding and research priorities.

**Thank you for your time and expertise!**

Click "Next" to begin.

---

## Category Selection Example

You selected the following area:

**Ethics:** Work on AI ethics includes developing methods for integrating ethical considerations into training, evaluation, and decision-making processes, as well as techniques for mitigating harmful outputs and ensuring cultural and long-term ethical consistency.

This area has the following sub-areas:

- Ethics-aware training and fine-tuning
- Ethical decision-making frameworks
- Mitigating harmful outputs
- Cultural sensitivity and contextual awareness
- Long-term ethical consistency

Please answer several questions about each sub-area on the following pages.

Feel free to skip questions or sub-areas that you are unfamiliar with.

## Sub-Area Evaluation Example

**Ethics-aware training and fine-tuning:** Research on learning from imperfect ethical datasets, applying ethics-aware data curation methods, and incorporating collective ethical principles into model design.

**Importance:** Resolving the core challenges of this sub-area and implementing the resulting solutions would significantly reduce the risk of severe harm (loss of >100 lives or >$10 billion in economic impact) from AI.

| ◯ | ◯ | ◯ | ◯ | ◯ | ◯ |
|---|---|---|---|---|---|
| Strongly Disagree | Disagree | Neutral | Agree | Strongly Agree | Don't Know |

**Tractability:** An additional targeted investment of approximately $10 million over the next two years would lead to significant, measurable advancements in addressing this sub-area's underlying challenges.

| ◯ | ◯ | ◯ | ◯ | ◯ | ◯ |
|---|---|---|---|---|---|
| Strongly Disagree | Disagree | Neutral | Agree | Strongly Agree | Don't Know |

What metrics or indicators would best measure progress in this area? Do you have any qualitative feedback or suggestions for improving work in this area?

---

## Category Feedback Example

Please answer one or more of the following questions about the high-level area: **Ethics**

**High-value challenge:** How would you phrase the single most impactful challenge in this area that could benefit from further funding? This can be a challenge mentioned previously in the survey, or a new one.

**Missing sub-areas:** Are there any other important sub-areas not covered previously?

**Key obstacles:** What major obstacles (technical, organizational, or regulatory) do you see as the main barriers to progress in this area?

---

## Demographic Questions

What is your primary area of expertise?

- Computer Science/Engineering
- Machine Learning/Artificial Intelligence
- Social Sciences (e.g. Economics)
- Humanities/Philosophy/Ethics
- Public Policy/Governance
- Other, please specify

What best describes your current position and background?

- Currently enrolled PhD Student (≥1 year into program)
- Postdoctoral Researcher/Fellow
- Faculty/Professor
- Industry professional with ≥3 years relevant R&D experience
- Policymaker or other professional with equivalent experience
- Other, please specify

How many years have you spent conducting research or working in your primary area of expertise (including postgraduate education and professional roles)?

- <3 years
- 3-5 years
- 5-10 years
- >10 years

What type of organization are you primarily affiliated with?

- Academic Institution
- Industry
- Government
- Non-Profit Organization
- Independent Researcher/Consultant
- Other, please specify

# Appendix C: Survey Results

This section includes the full table of survey results, followed by additional data on the sub-areas of highest and lowest consensus among respondents.

## Full Survey Results

This table presents results for the questions of *importance* and *tractability.* In the survey, we defined these as agreement with the following statements on a 5-point Likert scale:

- **Importance:** Resolving the core challenges of this sub-area and implementing the resulting solutions would significantly reduce the risk of severe harm (loss of >100 lives or >$10 billion in economic impact) from AI.
- **Tractability:** An additional targeted investment of approximately $10 million over the next two years would lead to significant, measurable advancements in addressing this sub-area's underlying challenges.

For each sub-area with sufficient responses (n>2 for both questions), we calculated mean scores for *importance* and *tractability*. We then computed a "promise score" for each sub-area by multiplying these unrounded means, reflecting a prioritization of areas that excel in both dimensions. Note that table values for *importance* and *tractability* are means rounded to two decimal places. Promise scores were calculated from unrounded means and then also rounded. This may result in minor apparent discrepancies if recalculating *promise* from the rounded table values.

For each area, we also provide the number of respondents (n) for both *importance* and *tractability*, as in some cases participants answered one and not the other.

For the full taxonomy with descriptions of each area, see Appendix A.

| | Research sub-area | Importance | (n) | Tractability | (n) | Promise |
|---|---|---|---|---|---|---|
| 1 | Emergence and task-specific scaling patterns | 5 | (3) | 4.25 | (4) | **21.25** |
| 2 | CBRN (Chemical, Biological, Radiological, and Nuclear) evaluations | 4.67 | (3) | 4.33 | (3) | **20.22** |
| 3 | Evaluating deception, scheming, situational awareness, and persuasion | 4.75 | (4) | 4.25 | (4) | **20.19** |
| 4 | Oversight and monitoring of LLM-agents | 4.67 | (9) | 4.22 | (9) | **19.7** |
| 5 | Cyber evaluations | 4.5 | (4) | 4.25 | (4) | **19.13** |
| 6 | Detecting and addressing previously unmeasured or latent capabilities | 4.38 | (16) | 4.25 | (16) | **18.59** |
| 7 | Multi-agent metrics and evaluations | 4.43 | (7) | 4.14 | (7) | **18.35** |

| | Research sub-area | Importance | (n) | Tractability | (n) | Promise |
|---|---|---|---|---|---|---|
| 8 | Multi-agent security | 4.57 | (7) | 4 | (6) | **18.29** |
| 9 | Quantifying cyber threats from advanced capabilities | 4.25 | (4) | 4.25 | (4) | **18.06** |
| 10 | Manage conflicts between different values | 4.5 | (4) | 4 | (4) | **18** |
| 11 | Safety and emergent functionality in multi-agent interactions | 4.57 | (7) | 3.86 | (7) | **17.63** |
| 12 | Mechanistic understanding and limits of LLM reasoning | 5 | (3) | 3.5 | (4) | **17.5** |
| 13 | Validating and applying interpretability methods | 4.38 | (8) | 4 | (7) | **17.5** |
| 14 | Evaluation methodology and metrics | 4.14 | (14) | 4.2 | (15) | **17.4** |
| 15 | Control mechanisms for untrusted models | 4 | (4) | 4.25 | (4) | **17** |
| 16 | Transparency, information asymmetries, and communication protocols | 4 | (5) | 4.25 | (4) | **17** |
| 17 | Detecting and preventing collusion and emergent collective behaviour | 4.43 | (7) | 3.8 | (5) | **16.83** |
| 18 | Network effects and destabilizing dynamics in agent ecosystems | 4.33 | (6) | 3.83 | (6) | **16.61** |
| 19 | Pluralistic value alignment | 4 | (4) | 4 | (4) | **16** |
| 20 | Evaluating tool affordances for LLM-agents | 4.11 | (9) | 3.89 | (9) | **15.99** |
| 21 | Develop more robust evaluations for which values an LLM encodes | 3.75 | (4) | 4.25 | (4) | **15.94** |
| 22 | Studying misalignment through simplified model organisms | 4.07 | (14) | 3.86 | (14) | **15.7** |
| 23 | Confidential computing and environment isolation | 4.57 | (7) | 3.43 | (7) | **15.67** |
| 24 | Theoretical foundations for evaluation | 4.07 | (15) | 3.81 | (16) | **15.5** |
| 25 | Improving evaluation robustness | 3.8 | (15) | 4.07 | (15) | **15.45** |
| 26 | Robustness to underspecification | 4.17 | (6) | 3.67 | (6) | **15.28** |
| 27 | Pretraining alterations to improve interpretability | 4 | (5) | 3.8 | (5) | **15.2** |
| 28 | Understanding how fine-tuning changes a pretrained model | 4 | (5) | 3.75 | (4) | **15** |
| 29 | Transparency | 4.13 | (8) | 3.63 | (8) | **14.95** |
| 30 | Foundational research on operationalizing values in LLMs | 4.25 | (4) | 3.5 | (4) | **14.88** |
| 31 | Weight security and key management | 4.5 | (4) | 3.25 | (4) | **14.63** |
| 32 | Debate | 4.1 | (10) | 3.5 | (10) | **14.35** |

| | Research sub-area | Importance | (n) | Tractability | (n) | Promise |
|---|---|---|---|---|---|---|
| 33 | Continuous monitoring, advanced threat detection, and incident response | 3.86 | (7) | 3.71 | (7) | **14.33** |
| 34 | Datacenter security | 4.14 | (7) | 3.43 | (7) | **14.2** |
| 35 | Uncertainty quantification | 3.75 | (4) | 3.75 | (4) | **14.06** |
| 36 | Defending against poisoning and backdoors | 3.75 | (4) | 3.75 | (4) | **14.06** |
| 37 | Justify value choices for alignment | 4 | (4) | 3.5 | (4) | **14** |
| 38 | Scalable data auditing, filtering, and Pretraining with Human Feedback (PHF) | 4 | (4) | 3.5 | (4) | **14** |
| 39 | Eliciting Latent Knowledge (ELK) | 4.25 | (8) | 3.25 | (8) | **13.81** |
| 40 | Explainability | 4 | (7) | 3.43 | (7) | **13.71** |
| 41 | Supply chain integrity and secure development | 4.57 | (7) | 3 | (6) | **13.71** |
| 42 | Develop output-based adversarial training techniques for more robust alignment | 3.8 | (5) | 3.6 | (5) | **13.68** |
| 43 | Better elicitation mechanisms from humans | 3.6 | (10) | 3.67 | (9) | **13.2** |
| 44 | Access control and interface hardening | 4.75 | (4) | 2.75 | (4) | **13.06** |
| 45 | Building verifiable and robust AI architectures | 3.92 | (12) | 3.33 | (12) | **13.06** |
| 46 | Lifelong learning and goal-directedness in LLM agents | 3.67 | (9) | 3.56 | (9) | **13.04** |
| 47 | Iterated Distillation and Amplification (IDA) | 4 | (9) | 3.22 | (9) | **12.89** |
| 48 | Tamper-evidence and tamper-proofing | 3.57 | (7) | 3.57 | (7) | **12.76** |
| 49 | Hardware-integrated monitoring and verification | 3.57 | (7) | 3.5 | (6) | **12.5** |
| 50 | Developmental interpretability | 3.75 | (8) | 3.29 | (7) | **12.32** |
| 51 | Defending against jailbreaks and prompt injections | 3.75 | (4) | 3.25 | (4) | **12.19** |
| 52 | Detecting modified models or poisoned data | 4 | (4) | 3 | (4) | **12** |
| 53 | Decision theory and rational agency | 3.7 | (10) | 3.2 | (10) | **11.84** |
| 54 | Formal verification of AI systems | 3.82 | (11) | 3.09 | (11) | **11.8** |
| 55 | Interpretability foundations | 3.38 | (8) | 3.43 | (7) | **11.57** |
| 56 | Peer incentivisation and automated mechanism design | 3 | (4) | 3.75 | (4) | **11.25** |
| 57 | Human-AI Interaction and collaboration | 3.2 | (5) | 3.5 | (4) | **11.2** |
| 58 | Preventing model self-exfiltration | 4 | (4) | 2.75 | (4) | **11** |
| 59 | Embedded agency | 3.55 | (11) | 3.09 | (11) | **10.96** |
| 60 | Feature and circuit analysis | 3.43 | (7) | 3.17 | (6) | **10.86** |

| | Research sub-area | Importance | (n) | Tractability | (n) | Promise |
|---|---|---|---|---|---|---|
| **61** | Specialized chips to compute encrypted data | 3.57 | (7) | 3 | (7) | **10.71** |
| **62** | Limiting models' ability to perform harmful tasks | 3.6 | (5) | 2.8 | (5) | **10.08** |
| **63** | Recursive Reward Modeling | 3.22 | (9) | 3.13 | (8) | **10.07** |
| **64** | Mechanism design and multi-agent communication | 2.8 | (5) | 3.5 | (4) | **9.8** |
| **65** | Adversarial robustness to perturbations | 3.25 | (4) | 3 | (4) | **9.75** |
| **66** | Retrieval-augmented pre-training | 2.6 | (5) | 3.75 | (4) | **9.75** |
| **67** | Scalable techniques for targeted modifications of LLM behavior | 3 | (4) | 3 | (5) | **9** |
| **68** | Theoretical foundations of deep learning | 2.67 | (3) | 3.33 | (3) | **8.89** |
| **69** | Reinforcement Learning from AI Feedback (RLAIF) | 3.1 | (10) | 2.8 | (10) | **8.68** |
| **70** | Limits of Transformers | 2.33 | (3) | 3.67 | (3) | **8.56** |
| **71** | Causal incentives | 2.9 | (10) | 2.9 | (10) | **8.41** |
| **72** | Control theory applications in AI safety | 3.09 | (11) | 2.6 | (10) | **8.04** |
| **73** | Model robustness and oracle protection | 3 | (4) | 2.67 | (3) | **8** |
| **74** | Double descent and overparameterization | 2 | (3) | 3.33 | (3) | **6.67** |
| **75** | Implicit bias of optimization algorithms | 2.67 | (3) | 2.33 | (3) | **6.22** |
| **76** | Optimization and loss landscape analysis | 2 | (3) | 2.33 | (3) | **4.67** |

## Consensus Analysis

Understanding the level of agreement among experts adds critical context to the average scores provided above. Low variance in ratings indicates stronger consensus, while high variance reflects significant disagreement about an area's importance or tractability.

### Areas with Strongest Expert Consensus

Our analysis identified several research areas with remarkably low variance in expert assessments, indicating strong agreement about their value:

| | Research sub-area | Importance Variance | (n) | Tractability Variance | (n) | Combined Variance |
|---|---|---|---|---|---|---|
| 1 | Oversight and monitoring of LLM-agents | 0.50 | (9) | 0.44 | (9) | **0.47** |
| 2 | Evaluation methodology and metrics | 0.44 | (14) | 0.46 | (15) | **0.45** |
| 3 | Validating and applying interpretability methods | 0.55 | (8) | 0.33 | (7) | **0.44** |
| 4 | Multi-agent security | 0.29 | (7) | 0.40 | (6) | **0.34** |
| 5 | Detecting and addressing previously unmeasured or latent capabilities | 0.38 | (16) | 0.20 | (16) | **0.29** |

### Areas with Significant Expert Disagreement

Conversely, several research areas showed high variance in expert assessments, indicating significant disagreement about their value:

| | Research sub-area | Importance Variance | (n) | Tractability Variance | (n) | Combined Variance |
|---|---|---|---|---|---|---|
| 1 | Building verifiable and robust AI architectures | 2.63 | (12) | 2.97 | (12) | **2.80** |
| 2 | Feature and circuit analysis | 2.95 | (7) | 2.17 | (6) | **2.56** |
| 3 | Datacenter security | 2.14 | (7) | 2.29 | (7) | **2.21** |
| 4 | Limiting models' ability to perform harmful tasks | 1.30 | (5) | 2.70 | (5) | **2.00** |
| 5 | Tamper-evidence and tamper-proofing | 1.95 | (7) | 1.95 | (7) | **1.95** |
| 6 | Hardware-integrated monitoring and verification | 1.95 | (7) | 1.90 | (6) | **1.93** |
| 7 | Interpretability foundations | 2.55 | (8) | 1.29 | (7) | **1.92** |
| 8 | Causal incentives | 1.43 | (10) | 2.32 | (10) | **1.88** |
| 9 | Robustness to underspecification | 1.77 | (6) | 1.87 | (6) | **1.82** |
| 10 | Specialized chips to compute encrypted data | 2.29 | (7) | 1.33 | (7) | **1.81** |

# Appendix D: Sub-areas excluded due to insufficient response

Sub-areas that received two or fewer ratings of either *importance* or *tractability* were excluded from analysis due to insufficient data for meaningful statistical interpretation.

| Category | Research sub-area |
|---|---|
| **Understanding in-context learning, reasoning, and scaling behavior** (4 of 7 sub-areas excluded) | Mechanistic understanding of In-Context Learning |
| | Influences on ICL behavior and performance |
| | Theoretical and representational aspects of scaling |
| | Impact of scaling and training on reasoning capabilities |
| **Domain-specific AI evaluation design** (1 of 4 sub-areas excluded) | Automated AI R&D evaluations |
| **Fairness** (6 of 6 sub-areas excluded) | Fairness under dynamic and constrained data scenarios |
| | Fair representation and participation in AI systems |
| | Bias detection, quantification, and mitigation techniques |
| | Fairness in multilingual, cross-cultural, and multimodal contexts |
| | Intersectional fairness and complex group structures |
| | Reconciling multiple fairness definitions and normative trade-offs |
| **Accountability** (4 of 4 sub-areas excluded) | Auditing mechanisms |
| | Methods for detecting and attributing LLM outputs |
| | Regulatory compliance automation |
| | Methods for mitigating power concentration in AI |
| **Ethics** (5 of 5 sub-areas excluded) | Ethics-aware training and fine-tuning |
| | Ethical decision-making frameworks |
| | Mitigating harmful outputs |
| | Cultural sensitivity and contextual awareness |
| | Long-term ethical consistency |
| **Privacy** (5 of 5 sub-areas excluded) | Identifying emergent privacy risks in new paradigms |
| | Research on inferring sensitive information from accumulation of innocuous data |
| | Privacy challenges in complex data scenarios |
| | Privacy modeling frameworks |
| | Data encryption tools for model inputs and outputs |
| **Research on safety in non-LLM systems** (4 of 4 sub-areas excluded) | Safe reinforcement learning for non-LLM systems |
| | Robotics and embodied AI safety |
| | Adversarial robustness in vision and perception systems |
| | Whole-brain emulation |

# Appendix E: Demographics

**Current Position and Background**

This describes the type of work that participants currently do–such as research, professorship, policymaking, and so on. About half of participants were in academic positions at the time of the survey, while over a quarter were working in industry research positions.

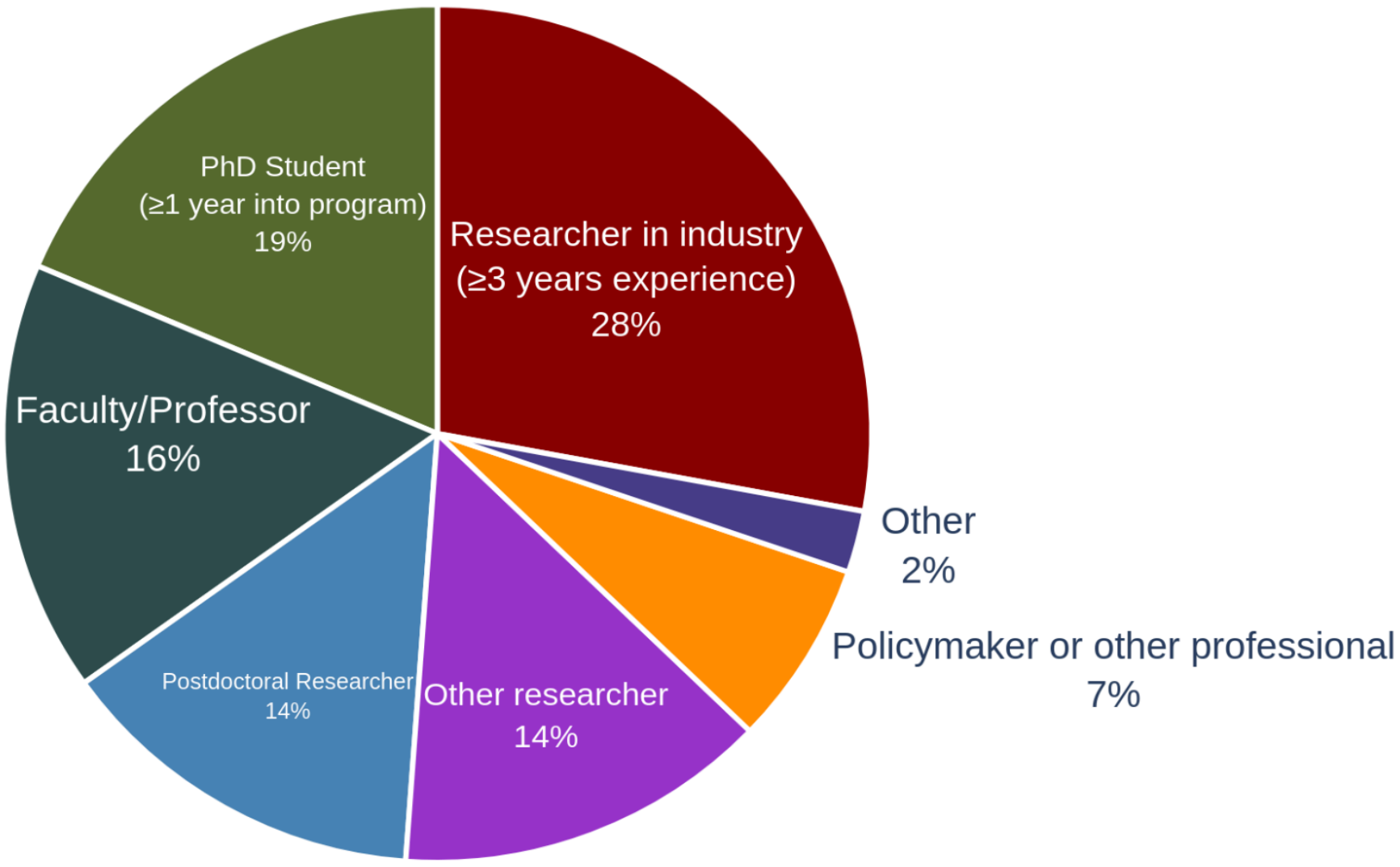

**Primary Organizational Affiliation**

This describes the type of organization that participants work in, such as academic institutions, non-profits, industry, and so on. Academic and non-profit affiliations dominated the respondent pool, with industry affiliations as another notable category.

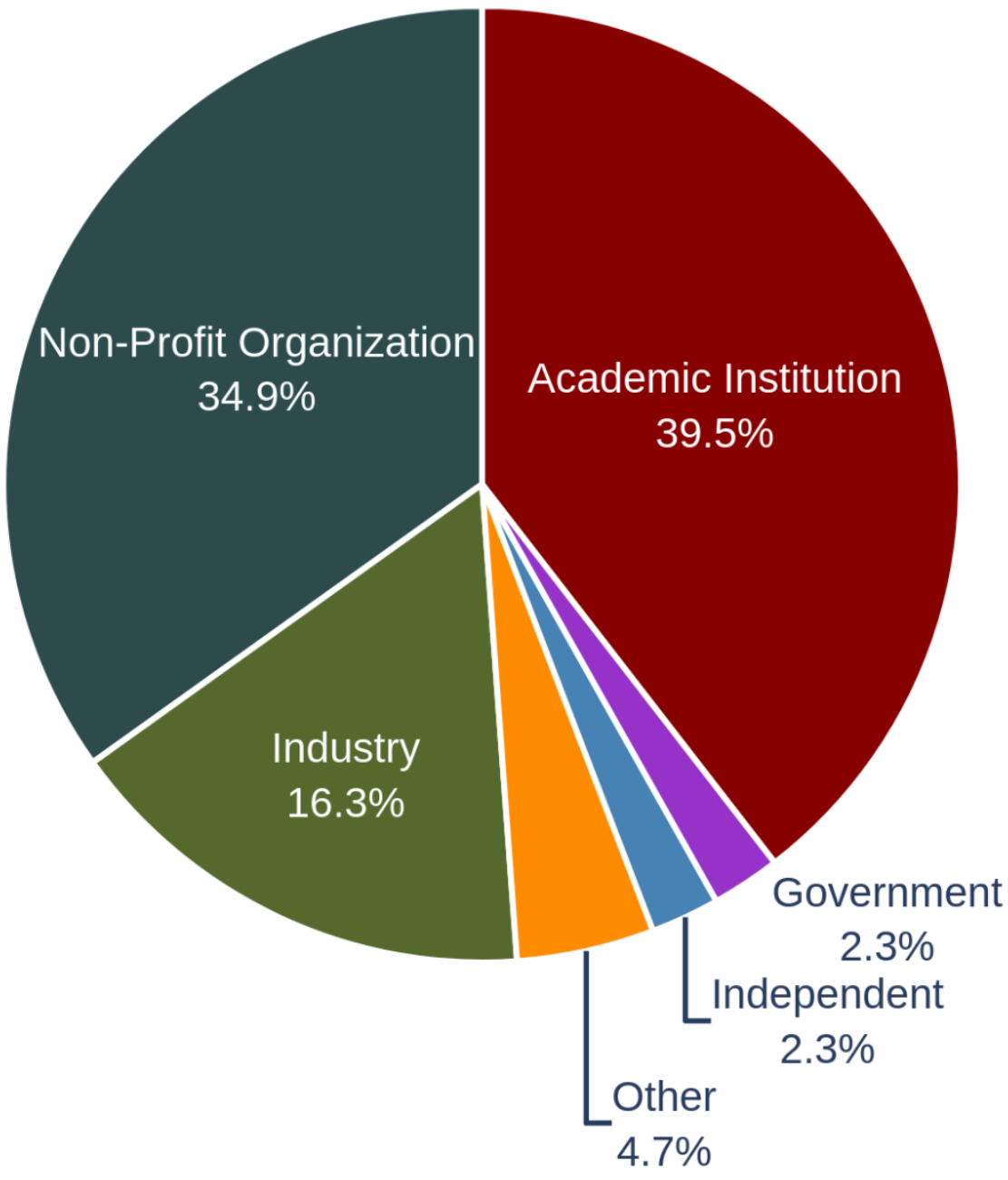

### Primary Area of Expertise

The primary areas of expertise among participants leaned strongly toward machine learning and computer science, which reflects the technical nature of the survey.

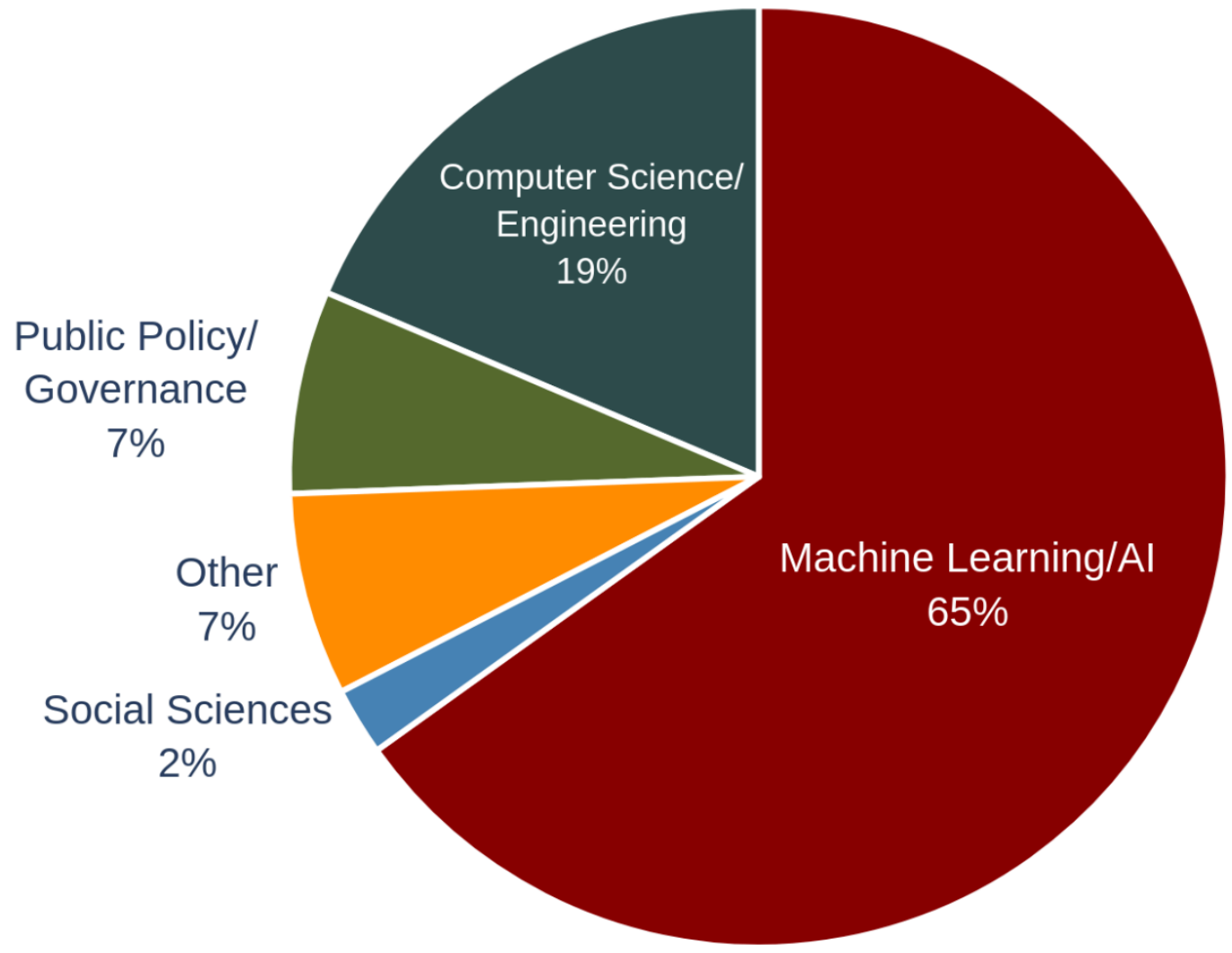

### Years Spent in Primary Area of Expertise

The modal experience level of participants was 5-10 years spent in their primary area of expertise, followed by 3-5 years, 10+ years, and <3 years.

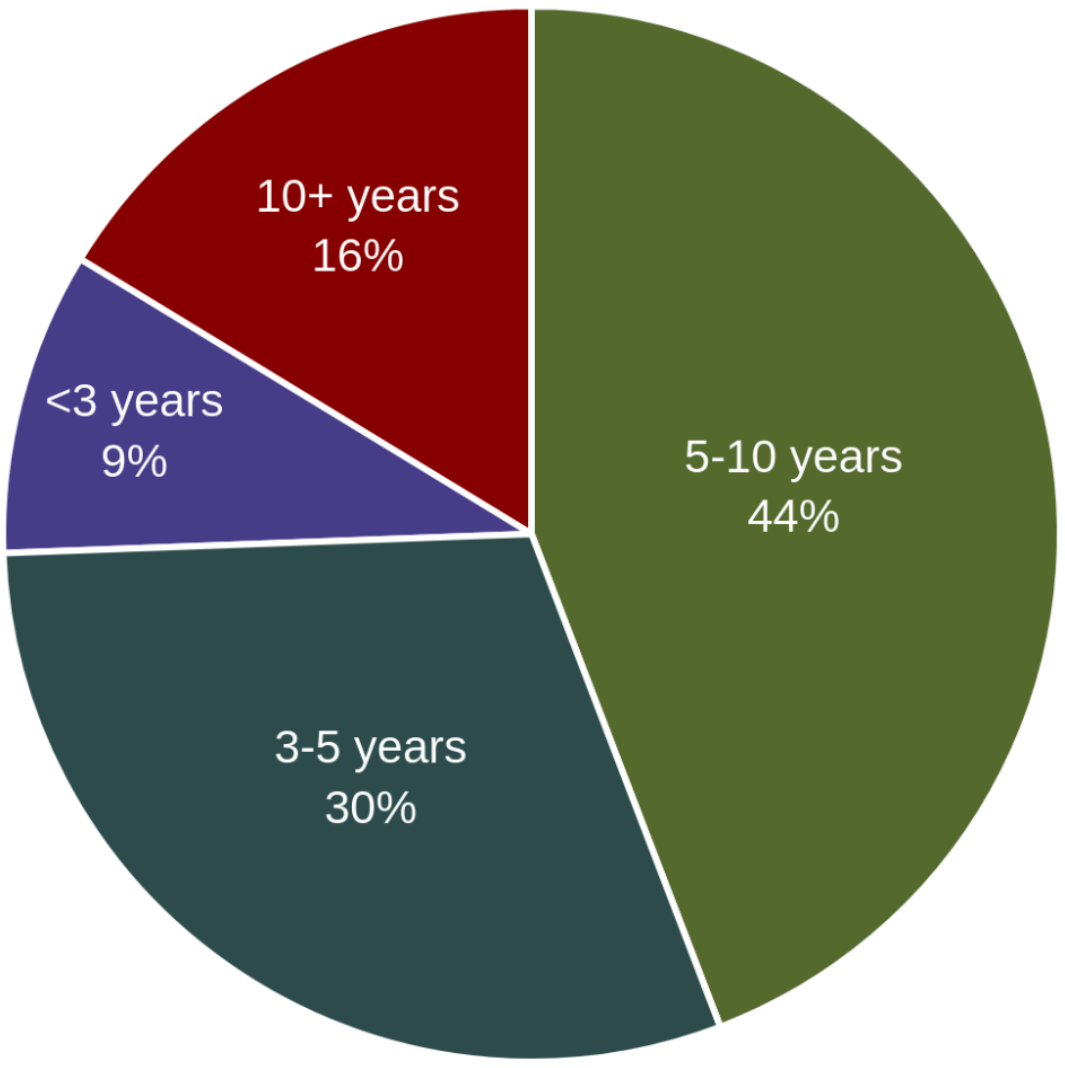